\PassOptionsToPackage{dvipsnames, table, x11names}{xcolor}
\documentclass[sigconf]{acmart}

\AtBeginDocument{%
  }

\renewcommand\footnotetextcopyrightpermission[1]{}
\setcopyright{acmlicensed}
\copyrightyear{2018}
\acmYear{2018}
\acmDOI{XXXXXXX.XXXXXXX}

\acmConference[XXX '25]{XXX}{XXXX--XXXX}{XX, XX}
\acmISBN{978-1-4503-XXXX-X/18/06}

\acmSubmissionID{xxx}

\newcommand{\up}[1]{\textcolor{OliveGreen}{\scriptsize \ $\uparrow${#1}}}
\newcommand{\downbad}[1]{\textcolor{Maroon}{\scriptsize \ $\downarrow${#1}}}

\usepackage{graphicx}
\usepackage{xcolor}
\usepackage{multirow}
\usepackage{rotating}
\usepackage{multicol}
\usepackage{booktabs}
\usepackage{amsfonts}

\usepackage{amssymb}
\usepackage{pifont}
\usepackage{cleveref}
\usepackage{enumitem}
\Crefname{figure}{Figures}{Figures}
\usepackage{nicefrac}  
\usepackage{xcolor}  
\usepackage{subfigure}
\usepackage{makecell}
\usepackage{xspace}
\usepackage[dvipsnames, table, x11names]{xcolor}
\usepackage{graphicx}
\usepackage{caption}
\usepackage{subcaption}
\usepackage[dvipsnames,table]{xcolor}
\definecolor{myblue}{RGB}{245,245,250}
\newcommand{\name}{\textit{HyperG}\xspace}
\newcommand{\etal}{\emph{et al.}\xspace}
\newcommand{\eg}{\emph{e.g., }\xspace}
\newcommand{\ie}{\emph{i.e., }\xspace}

\begin{document}

\title{Debiasing Sequential Recommendation with Time-aware Inverse Propensity Scoring}

\author{Sirui Huang}
\orcid{0000-0003-1206-2260}
\email{sirui.huang@connect.polyu.hk}
\affiliation{%
  \institution{Hong Kong Polytechinic University}
  \city{Hong Kong SAR}
  \country{China}
}
\affiliation{%
  \institution{University of Technology Sydney}
  \city{Sydney}
  \country{Australia}
}

\author{Jing Long}
\orcid{0000-0003-0848-2158}
\email{jing.long@uts.edu.au}
\affiliation{%
  \institution{University of Technology Sydney}
  \city{Sydney}
  \country{Australia}
}

\author{Qian Li}
\orcid{0000-0002-8308-9551}
\affiliation{
  \institution{Curtin University}
  \city{Perth}
  \country{Australia}
  }

\author{Guandong Xu}
\orcid{0000-0003-4493-6663}
\affiliation{%
  \institution{Education University of Hong Kong}
  \city{Hong Kong SAR}
  \country{China}
}
\affiliation{%
  \institution{University of Technology Sydney}
  \city{Sydney}
  \country{Australia}
}

\author{Qing Li}
\orcid{0000-0003-3370-471X}
\email{qing-prof.li@polyu.edu.hk}
\authornotemark[2]
\affiliation{%
  \institution{Hong Kong Polytechinic University}
  \city{Hong Kong SAR}
  \country{China}
}

\renewcommand{\shortauthors}{Huang et al.}

\begin{abstract}
Sequential Recommendation (SR) predicts users’ next interactions by modeling the temporal order of their historical behaviors. Existing approaches, including traditional sequential models and generative recommenders, achieve strong performance but primarily rely on explicit interactions such as clicks or purchases while overlooking item exposures. This ignorance introduces selection bias, where exposed but unclicked items are misinterpreted as disinterest, and exposure bias, where unexposed items are treated as irrelevant. Effectively addressing these biases requires distinguishing between items that were ``not exposed'' and those that were ``not of interest'', which cannot be reliably inferred from correlations in historical data. Counterfactual reasoning provides a natural solution by estimating user preferences under hypothetical exposure, and Inverse Propensity Scoring (IPS) is a common tool for such estimation. However, conventional IPS methods are static and fail to capture the sequential dependencies and temporal dynamics of user behavior. To overcome these limitations, we propose Time aware Inverse Propensity Scoring (TIPS). Unlike traditional static IPS, TIPS effectively accounts for sequential dependencies and temporal dynamics, thereby capturing user preferences more accurately. Extensive experiments show that TIPS consistently enhances recommendation performance as a plug-in for various sequential recommenders. Our code will be publicly available upon acceptance.

\vspace{-2mm}
\end{abstract}




\received{20 February 2007}
\received[revised]{12 March 2009}
\received[accepted]{5 June 2009}

\maketitle
\section{Introduction}\label{sec:intro}
Sequential Recommendation (SR) aims to predict users' next interactions by modeling the temporal order of historical interactions. Rather than treating each interaction as an independent event, SR models capture evolving preferences by leveraging sequence-aware models, allowing the system to infer both short-term and long-term interests, resulting in more accurate recommendations than methods based solely on static user profiles.

\begin{figure}[t!]
    \centering
    \includegraphics[width=0.8\columnwidth]{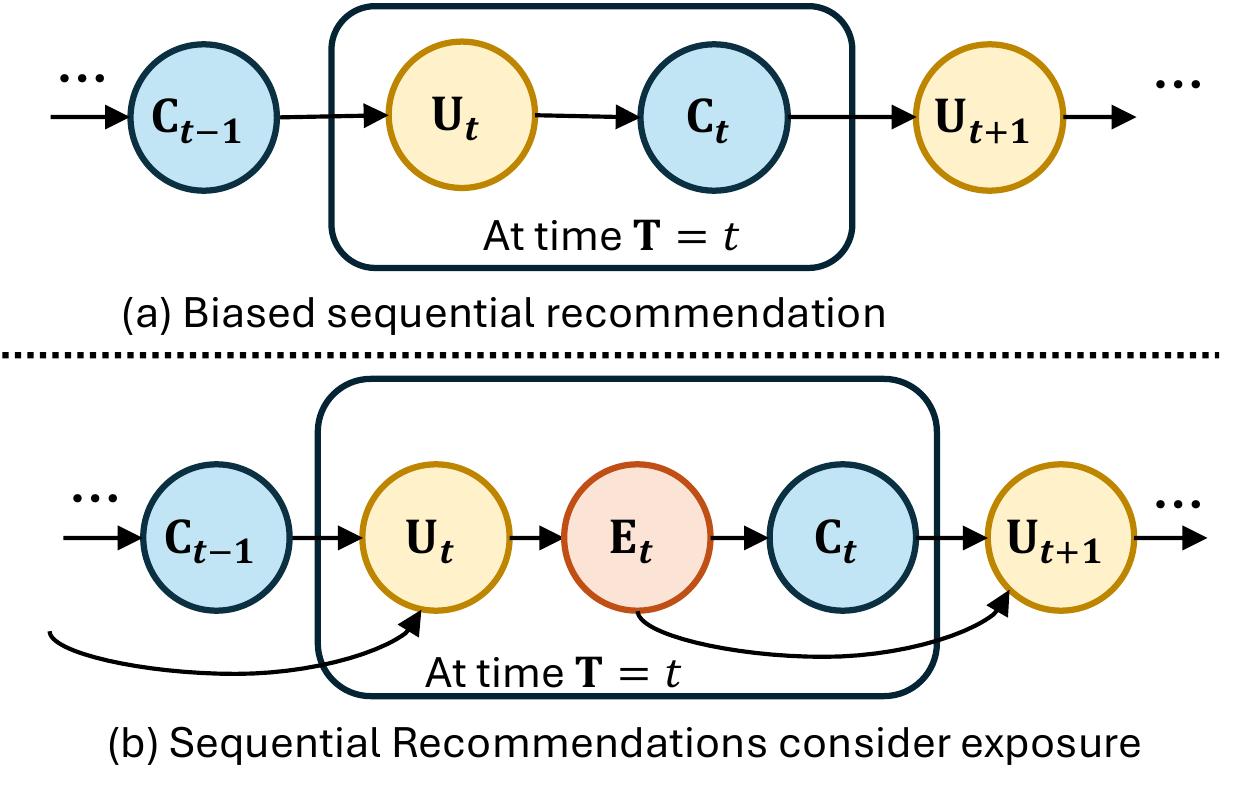}
    \vspace{-5mm}
    \caption{(a) An example of a biased SR model that learns user preference solely from interactions. (b) Example of a system that considers exposure. Here, $\mathbf{C}_t$ denotes user interactions (\eg clicks) at time $t$, $\mathbf{U}_t$ represents the user’s preference at time $t$, and $\mathbf{E}_t$ is the recommendation list containing all items exposed to the user at time $t$.}
    \label{fig:toy_example}
    \vspace{-5mm}
\end{figure}

These sequential recommenders fall broadly into two categories.
First, traditional sequential models, such as Recurrent Neural Networks (RNNs)~\cite{gru4rec} and transformer-like architectures~\cite{vaswani2017attention,sun2019bert4rec,liu2024mamba4rec}, are employed to calculate the probabilities of candidate items based on explicit feedback in historical interaction sequences to rank the items in recommendation lists~\cite{sun2019bert4rec,gru4rec,sasrec,10.1145/3616855.3635760}. 
For example, Yuan \etal use a transformer-based architecture to capture multiple types of positive user feedback, including purchases, clicks, and adding to cart, in order to enrich user sequential pattern modeling~\cite{sun2019bert4rec}. Second, generative recommendation systems, inspired by advances in natural language processing, aim to learn the distribution of historical interactions and directly generate the next interacted item~\cite{diffurec,pdrec,tdiffrec,ddrm,10.1007/978-981-97-2262-4_13}. For example, Li \etal leverages the distributions within diffusion models to dynamically represent the diverse interests of a user and different features of an item~\cite{diffurec}. In both of the categories, most current sequential recommenders rely solely on explicit interactions. While some methods attempt to incorporate contextual information to adjust for bias, most existing approaches primarily depend on explicit feedback signals (e.g., purchases or clicks) from users’ interaction histories to improve prediction accuracy. However, existing SR models risk amplifying biases by ignoring item exposures. As depicted in Fig. \ref{fig:toy_example}, a SR model that relies solely on interactions (\eg clicks) while ignoring item exposures learns a biased view of user preference, as it only models user feedback on a limited set of pre-interacted items (\ie existing works only modeling $\mathbf{C}\rightarrow\mathbf{U}$ but ignore $\mathbf{E}\rightarrow\mathbf{C}\rightarrow\mathbf{U}$ and $\mathbf{E}\rightarrow\mathbf{U}$).

The absence of item exposure mainly causes two biases in SRs: i) \textbf{exposure bias}, where items that were never exposed are implicitly treated as if the user showed no interest, ii) \textbf{selection bias}, where items not interacted are treated as no interest, though the lack of interaction may stem from display strategy or context. Effectively addressing these two biases requires distinguishing between items that were ``not exposed'' and those that were ``not of interest'', as missing interactions in training data may arise from either. However, in the absence of exposure logs, this distinction cannot be reliably inferred from correlations across users' historical data. To this end, counterfactual reasoning is conditioned on the hypothetical situation where the item is exposed to the user\cite{10.1145/3726302.3730005}. Specifically, counterfactual reasoning distinguishes not exposed from no interest in addressing \textbf{exposure bias} by asking, ``\textit{What if this item had been exposed to the user?}'', while it tackles \textbf{selection bias} by asking, ``\textit{What if the user had shown interest given exposure?}'', thereby separating cases of exposed but no interest from potential interest. 
Existing counterfactual reasoning often relies on inverse propensity scoring (IPS), which reweights observed interactions by the inverse of their exposure probabilities\cite{liu2025csrec,yang2024debiasing}. However, IPS is inherently static: it assigns weights to each interaction independently using only contextual signals (\eg user attributes or item features), without modeling temporal dependence or cross-interaction dynamics. This static nature limits IPS’s ability to fully correct biases in SR, as these biases can accumulate over time. i) The first challenge is that static IPS ignores the sequential dependencies among items in the interaction sequences. In SR, a user’s next interaction depends not only on item features but also on past interactions, as user interests evolve over time and previous recommendations may influence subsequent actions. Simply reweighting interaction samples can disrupt the causal chain within user interaction sequences (e.g., a user who interacts with an iPhone 17 is more likely to subsequently interact with a phone case). ii) Second, while the IPS reweighting simulates fair exposure across all items, it does not reflect users’ genuine interests that often follow temporal trends (\eg a user who interacts with an iPhone 17 is more likely to be interested in the newly released iWatch). Because an item’s factual exposure probability can shift over time due to new releases, promotions, or trending popularity, current IPS methods may over-weight newly introduced items while under-weighting items that are currently trending. 
To accurately reweight user interactions, IPS should account for these sequential and temporal dynamics, aligning estimated exposure with the true user interests. 

To overcome the limitations of conventional IPS in sequential recommendation for handling selection and exposure biases, we propose a novel framework, Time-aware Inverse Propensity Scoring (\name), which incorporates time-aware information in a model-agnostic manner. It improves both traditional sequential models and generative models by more accurately modeling users' implicit feedback and capturing sequential and time-varying causal dependencies in interactions. Specifically, instead of calculating the IPS for each user-item interaction directly, we first generate two high-quality counterfactual interactions for each factual interaction observed in the training data. One counterfactual varies the time representation of the interaction, and the other perturbs the representation for the item itself. Combined with historical interactions, these two counterfactual interactions
help accurately model time-aware inverse propensity under the sequential and time-varying nature of user behavior. 
The main contributions of \name are summarized as follows:
\begin{itemize}[leftmargin=*]
    \item We estimate the item exposure distribution by constructing time-aware counterfactual examples in the absence of exposure logs.
    \item We mitigate exposure-related biases, including exposure and selection bias, and more accurately model user preferences by reweighting recommendations with time-aware propensities in the absence of exposure logs.
    \item We conduct extensive experiments on three backbones and compare \name with two categories of recommendation models to validate its effectiveness.
\end{itemize}

\section{Preliminary}
In this section, we first propose the structure causal model (SCM) (Fig. \ref{fig:scm}) to clarify the causal relationships in SR. Then, we introduce the traditional Inverse Propensity Scoring (IPS) method, followed by the definition of problem that this paper focuses on. 

Suppose a user $u\in\mathcal{U}$ interacts with an item $v\in\mathcal{V}$ at time $t\in\mathcal{T}$, where $\mathcal{U}$, $\mathcal{V}$, $\mathcal{T}$ denote the set of users, items, and all the interaction time periods, repectively. In the causal model (Fig. \ref{fig:scm}), the variable $\mathbf{U}$ denotes user preferences, the binary variables $\mathbf{E}\in\{0,1\}$ and $\mathbf{C}\in\{0,1\}$ represent whether an item is exposed to or clicked by user, respectively. 
In this paper, all equations are formulated for a single user $u$ for clarity. Here, $\mathbf{C}_v$ and $\mathbf{E}_v$ are binary variables indicating whether the item $v$ is clicked and exposed, respectively. For example, $\mathbf{C}_v=1$ and $\mathbf{E}_v=1$ denotes the item $v$ is exposed and interacted with by the user $u$, respectively. 

\subsection{Structural Causal Model for SR}
\begin{figure}[t!]
    \centering
    \includegraphics[width=0.8\columnwidth]{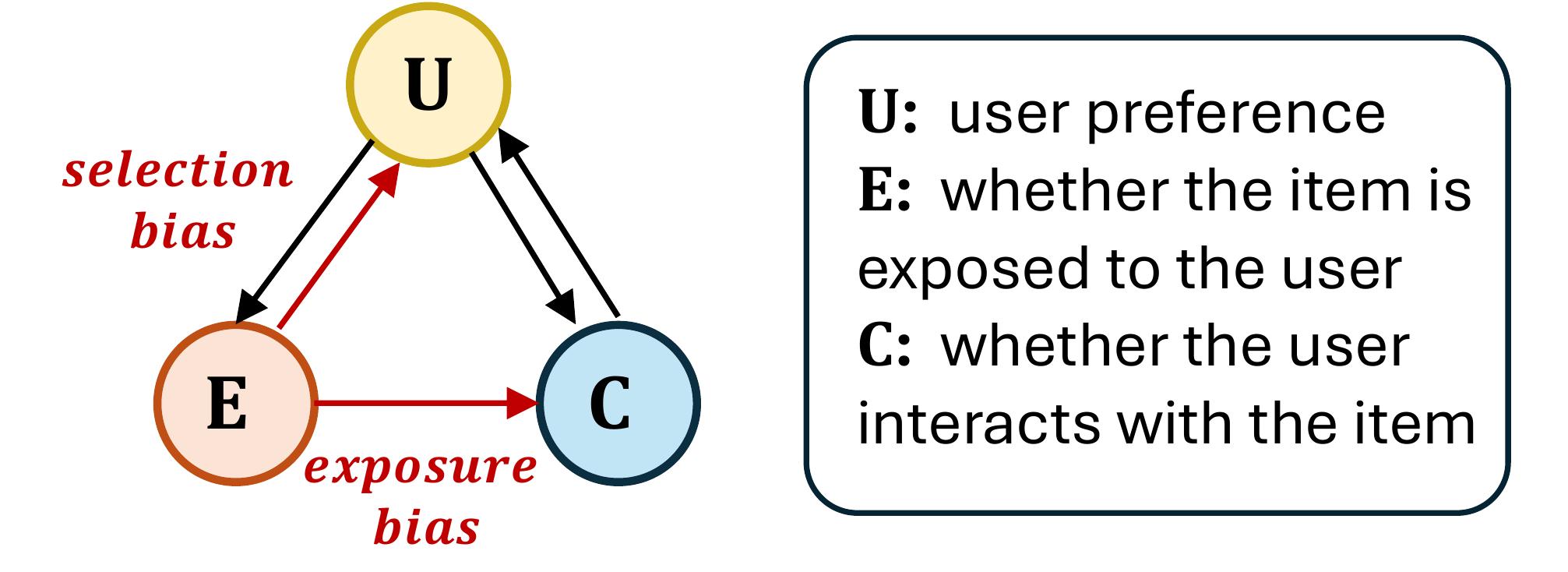}
    \vspace{-3mm}
    \caption{Structural Causal Model (SCM) of SR.}
    \label{fig:scm}
    \vspace{-5mm}
\end{figure}

Fig. \ref{fig:scm} illustrates the causal relationships in SRs. Different from Fig. \ref{fig:toy_example}, this Structural Causal Model (SCM) illustrates the exposure-related causal relationships under a fixed time, which allows for a clearer explanation. The directed arrow in Fig. \ref{fig:scm} indicates the information flow from the cause variable to the effect variable. Each edge in Fig. \ref{fig:scm} is described as follows.
\begin{itemize}[leftmargin=*]
    \item $\mathbf{U}\rightarrow\mathbf{E}$ represents that users' preference ($\mathbf{U}$) influence which items are exposed in their recommendation list ($\mathbf{E}$).
    \item $\mathbf{E}\rightarrow\mathbf{U}$ represents that, besides interactions data, the exposure of items in a recommendation list ($\mathbf{E}$) also reflects user's true preference ($\mathbf{U}$) (selection bias).
    \item $\mathbf{U}\rightarrow\mathbf{C}$ represents that users' preferences ($\mathbf{U}$) determines which items are interacted with (\eg clicked) ($\mathbf{C}$).
    \item $\mathbf{C}\rightarrow\mathbf{U}$ represents that user interactions ($\mathbf{C}$) provide information about user preferences ($\mathbf{U}$).
    \item $\mathbf{E}\rightarrow\mathbf{C}$ represents that interactions ($\mathbf{C}$) occur only among exposed items ($\mathbf{E}$) (exposure bias).
\end{itemize}

\textbf{Exposure bias} arises because recommenders only observe a subset of items that are exposed to the user ($\mathbf{E}=1$), while the user's preference toward unexposed items ($\mathbf{E}=0$) remains unobserved. As illustrated in Fig. \ref{fig:scm}, exposure $\mathbf{E}$ is a prerequisite for interaction $\mathbf{C}$. For example, when a user is only exposed to ten movies, the model cannot infer whether the user would have interacted with the unexposed movies other than these ten that never appeared in the recommendation list.  Consequently, the model can only estimate the interaction probability conditioned on exposure, \ie $P(\mathbf{C}|\mathbf{E}=1)$, rather than the modeling the full distribution over both exposed and unexposed items (\ie $P(\mathbf{C}|\mathbf{E}\in\{0,1\})$). Formally, the exposure bias can be formulated as follows:

\begin{equation}
     P(\mathbf{C}|\mathbf{E}=1)\neq P(\mathbf{C}|\mathbf{E}\in\{1,0\}), P(\mathbf{E}=1)<1
\end{equation}
where the left-hand side of the inequality represents the biased interaction distribution learned only from exposed items, while the right-hand side corresponds to the ideal interaction distribution conditioned on the exposure probability of items. These two distributions are intuitively not equal because the exposure probability of an item is generally less than 1, and the interactions $\mathbf{C}$ inherently depend on item exposure $\mathbf{E}$.

As illustrated in Fig. \ref{fig:scm}, the interaction variable $\mathbf{C}$ is determined by both the exposure $\mathbf{E}$ and the user preference $\mathbf{U}$. However, in the absence of exposure logs, most recommendation models rely only on the subset of items with $\mathbf{C}=1$, \ie interacted items. This introduces \textbf{selection bias}, since the model is deviated from the observed interactions with selected items. For example, a user may be exposed to ten movies but only click one of them. If the model only learns from the clicked movie, it wrongly assumes the user has no preference over the rest of nine movies.

\begin{equation}
    P(\mathbf{U}|\mathbf{C}=1, \mathbf{E}=1)\neq P(\mathbf{U}|\mathbf{E}=1)
\end{equation}

\noindent
where the left-hand side represents the conditional distribution of user preferences over observed interactions, while the right-hand side represents the ideal preference distribution over all exposed items. These two distributions differ due to the existence of items that are exposed but not interacted with.

In the absence of exposure logs, existing SR models leverages explicit user feedback mainly consider the paths $\mathbf{U}\rightarrow\mathbf{C}$, $\mathbf{C}\rightarrow\mathbf{U}$, and $\mathbf{U}\rightarrow\mathbf{E}$, while overlooking the need to address exposure bias (path $\mathbf{E}\rightarrow\mathbf{C}$) and selection bias (path $\mathbf{E}\rightarrow\mathbf{U}$).

\subsection{Inverse Propensity Scoring}\label{sec:ips}
Inverse Propensity Scoring (IPS) is a statistical method widely adopted to address the exposure and selection biases by reweighting user-item interactions~\cite{liu2025csrec,yang2024debiasing,zhang2024uncovering,wang2024causally,xu2024rethinking,liao2024modeling}. 
\subsubsection{Propensity Score}
IPS-based recommenders use propensity scores to reweight interactions and correct bias caused by unobserved exposures. Specifically, the propensity score models the probability that an item is exposed to a user and is used to reweight observed interactions toward an ideal distribution with uniform exposure. This reweighting prevents underestimation of user preferences for rarely exposed items. Formally, the propensity score for the interaction between item $v$ and user $u$ can be defined as:
\begin{equation}
    \pi(u,v) = P(\mathbf{E}_v=1|u)
\label{eq:pi}
\end{equation}

\noindent
where $\pi(\cdot)$ denotes the exposure distribution estimation function. Therefore, lower propensity scores are assigned to items that has higher probability to be exposed, \ie items with higher $P(\mathbf{E}=1)$.

\subsubsection{Using IPS to Address the Two Biases} 
The \textbf{exposure bias} in the recommendations arises because the recommenders treated the unexposed items as not interested. Suppose we have an unexposed item $v$. To adjust for exposure bias using IPS, consider the ``what-if'' question in Section \ref{sec:intro}: ``\textit{What if this item had been exposed to the user?}'' Our goal is to estimate the interaction probability over an ideal distribution where all items have an equal chance of being exposed. Most recommenders ignore the important fact that exposure is a prerequisite for interaction with an item (\ie the path $\mathbf{E} \rightarrow \mathbf{C}$). To address this ignorance, IPS-based methods reweight the recommendation score $g(u,v)$ for each observed user-item interaction by the inverse of the item's exposure probability.
Formally,

\begin{equation}
\sum_{(u,v) \in \mathcal{D}_{\text{train}}} \frac{g(u,v)}{\pi(u,v)} P(\mathbf{U}=u) \simeq \mathbb{E}_{\mathbf{U},\mathbf{C}}[g(u,v)] 
\label{eq:ips_exp}
\end{equation}

\begin{equation}
\mathbb{E}_{\mathbf{U},\mathbf{C}}[g(u,v)] 
= \sum_{u\in\mathcal{U}} P(\mathbf{U}=u) \sum_{v\in\mathcal{V}} P(\mathbf{C}_v=1 \mid u) g(u,v)
\label{eq:ideal_exp}
\end{equation}

\noindent
The left-hand side of Eq. (\ref{eq:ips_exp}) represents the expectation of the recommendation scores reweighted by IPS for each observed interaction, where $\pi(u,v)$ denotes the propensity score for the specific interaction between user $u$ and item $v$. The right-hand side of Eq. (\ref{eq:ips_exp}) represents the ideal expectation of recommendation scores, which indicate the interaction probabilities between users and items under the true user preference distribution $P(\mathbf{U}=u)$ defined in Eq. (\ref{eq:ideal_exp}).


The \textbf{selection bias} refers to recommendations models learns only from observed interactions of the subset of interacted items. We have an unobserved (\ie non-interacted) item $v$. To address selection bias using IPS, consider the aforementioned ``what-if'' question: ``\textit{What if user $u$ had shown interest, given the exposure of item $v$?}'' We aim at inferring the user's true preference $\mathbf{U}$ over an ideal distribution where all items have an equal chance of being exposed. Considering the absence of exposure logs, the models do not account for the exposure probability in user modeling (\ie path $\mathbf{E}\rightarrow\mathbf{U}$ in Fig. \ref{fig:scm}). Traditional IPS corrects this by applying the inverse of propensity score $\pi(\cdot)$ to the score computed by the recommendation model $g$.

\begin{equation}
\sum_{\substack{u \in \mathcal{U}, \mathbf{C}_v=1, \mathbf{E}_v=1}} 
\frac{g(u,v)}{\pi(u,v)}
\simeq 
\sum_{\substack{u \in \mathcal{U}}} 
P(\mathbf{U}=u)[g(u,v)]
\label{eq:ips_select}
\end{equation}

\noindent
The left-hand side of Eq. (\ref{eq:ips_select}) represents the expectation of the recommendation scores reweighted by IPS. The right-hand side represents the expectation of ideal recommendation scores, where $P(\mathbf{U}=u)$ is the preference of user $u$ in the true user preference distribution.

\subsection{Problem Definition}
In SRs, the exposure probabilities of items described by propensity scores $\pi(\cdot)$ as formulated in Eq. (\ref{eq:pi}) varies over time, as user interests and exposure strategies (\eg, new release, promotions) changes. Therefore, this paper addresses exposure and selection biases using IPS, while accounting for the time-varying nature of SRs. Specifically, our goal is to leverage a plug-in model $f_\varphi$ that learns the item exposure distribution $P(\mathbf{E}=1|u)$ from observed factual interactions for any recommendation models $g_\theta$. We estimates item exposure over time and dynamically reweights interactions with the inverse of time-aware exposure probability, approximating a counterfactual setting where all items share the same chance of being exposed. This allows inferring interaction probabilities while accounting for the sequential and temporal dynamics of item exposure.
For each $u\in\mathcal{U}$, $v\in\mathcal{V}$, and the interaction time $t\in\mathcal{T}$, there is an unbiased recommendation score $\hat{y}(\cdot)$:

\begin{equation}
    \hat{y}(u,v,t) \approx (f_\varphi\circ g_\theta)(u,v,t)
\end{equation}

\noindent
where $\circ$ denotes the functional composition of the plug-in model $f_\varphi$ for exposure estimation and  the recommendation model $g_\theta$, parameterized by $\varphi$ and $\theta$, respectively.


\section{Methodology}
\begin{figure*}[htbp]
    \centering
    \includegraphics[width=\textwidth]{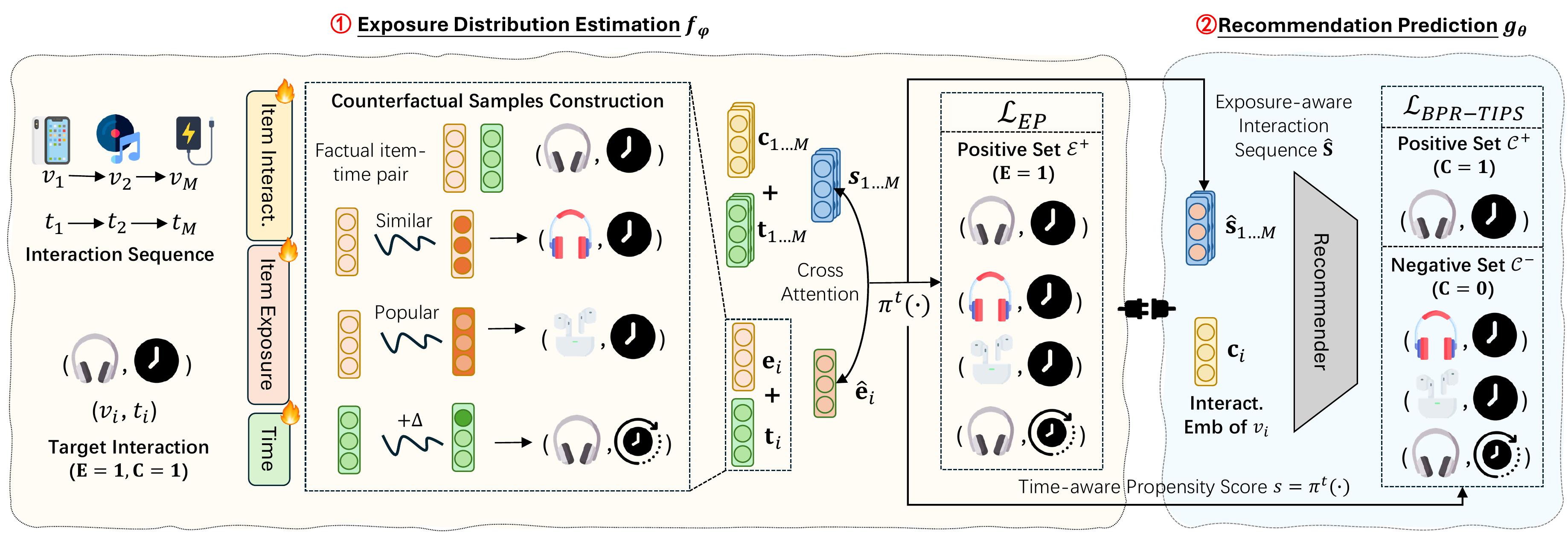}
    \vspace{-7mm}
    \caption{Overview of the Time-aware Inverse Propensity Scoring (TIPS) framework, designed as a plug-in for any SR model.}
    \label{fig:framework}
\end{figure*}

This paper focuses on using IPS to account for sequential and temporal causal relationships in SRs and addressing exposure and selection biases, which are two major biases arising from the absence of exposure data. We introduce time-aware inverse propensity scoring (TIPS), which incorporates temporal information into IPS to accurately capture the dynamics of SRs. Specifically, we propose a novel framework (Fig. \ref{fig:framework}) that integrates a plug-in exposure estimation model $f_\varphi$ with TIPS to optimize the recommender’s training objective. The model $f_\varphi$ estimates item exposure distributions by reasoning over counterfactual item–time pairs, which are then used to compute time-aware propensity scores for bias correction.


\subsection{Dual Encoding Strategy}
Before performing counterfactual reasoning, we encode both items and timestamps to disentangle static item semantics from the dynamics of user preferences and item exposures, enabling IPS to adjust toward an unbiased exposure distribution. This dual encoding involves maintaining two separate item embedding lookup tables and extracting normalized temporal features.

\subsubsection{Item Encoding}\label{sec:item_enc} Based on the structural causal model shown in Fig. \ref{fig:scm}, it is necessary to separately model the interaction distribution and exposure distribution for items. To this end, we maintain two hidden embedding matrices, $\mathbf{H}^{(C)}$ and $\mathbf{H}^{(E)}$. Each item $v_i \in \mathcal{V}$ is associated with two trainable embeddings, defined as follows.

\begin{equation}
\label{eq:item_emb}
\mathbf{c}_i = \operatorname{Lookup}(\mathbf{H}^{(C)},i),\ \mathbf{e}_i = \operatorname{Lookup}(\mathbf{H}^{(E)},i) 
\end{equation}

\noindent
where the embedding $\mathbf{H}^{(C)}\in\mathbb{R}^{|\mathcal{V}|\times d}$ captures interaction information, learned from historical user-item explicit feedback such as clicks, and encodes item semantics and collaborative information that reflect user preferences. The exposure embedding $\mathbf{H}^{(E)}\in\mathbb{R}^{|\mathcal{V}|\times d}$ represents the distribution of item exposure, involving factors such as popularity or promotional campaigns that influence whether an item is shown to a user.

The necessity of maintaining two separate lookup embeddings arises from two key considerations: i) \textit{Interaction and exposure reflect different user feedback.} Interacted items represent a user’s positive preferences, whereas exposed items, especially those not interacted with, may indicate disinterest. Both types of signals are valuable but inherently distinct. ii) \textit{Shared embeddings can amplify bias} in further inverse propensity scoring, which requires the propensity estimated (\ie exposure probability, the denominator) to be independent from the recommendation prediction (the numerator). Shared embeddings create dependency, which increases variance and bias in subsequent adjustment process.

\subsubsection{Time Embeddings}\label{sec:time_enc} The interaction time $t\in\mathcal{T}$ in the dataset are arranged sequentially in chronological order, with each $t_i$ representing the time of user $u$ interacting with item $v_i$. The gaps between two interactions matters because user preference changes. A commonly observed pattern is that more recent interactions better represent a user’s current preferences. For example, if a user watched a comedy yesterday and a documentary six months ago, the recommender should prioritize the recent comedy when recommending what to watch next. We compute the difference between the timestamp $t_i$ at position $i$ and the timestamp $t_{i-1}$ at position $i-1$, and then apply normalization. The normalized time interval is subsequently mapped into the embedding space with a multi-layer perceptron $\operatorname{MLP}(\cdot)$. Formally, $\mathbf{t}_i=\operatorname{MLP}(\operatorname{Norm}(t_i-t_{i-1}))$,
where $\operatorname{Norm}(\cdot)$ represents the normalization process, $\mathbf{t}_i\in\mathbb{R}^{|\mathcal{T}|\times d}$ is the embedding table of time.

\subsection{Counterfactual Samples Construction}
To estimate item exposure without exposure data, we construct counterfactual samples based on three assumptions that reflect common patterns in recommendation scenarios. For each interaction $(u, v_i, t_i)\in\mathcal{D}_{\text{train}}$, the item $v_i$ is positively exposed and clicked at time $t_i$, \ie, for user $u$, there is $\mathbf{E}=1, \mathbf{C}=1$. For each factual item–time pair $(v_i, t_i)$ in the interaction data $\mathcal{D}_{\text{train}}$, we assume three additional positive exposure item-time pairs where $\mathbf{E}=1, \mathbf{C}=0$:

i) \textbf{Similar Items $(v_i^{sim})$} are more likely to be exposed at the same time, as recommendation algorithms often promote related alternatives together~\cite{10.1145/3397271.3401177,10.1145/3568022,fan2022comprehensive}. Based on this assumption, we ask ``\textit{What if the user was shown a similar item?}'' To answer this counterfactual question, we retrieve similar items based on the exposure embedding table $\mathbf{H}^{(E)}$, which aims at learning the exposure distribution of items. The exposure embedding of the counterfactual similar item $v_i^{sim}$ can be formally expressed as: 
\begin{equation}
    \mathbf{e}_i^{sim} = \arg\max_{v_j \in \mathcal{V}\backslash v_i} \operatorname{sim}(\mathbf{e}_j, \mathbf{e}_i),\
\text{sim}(v_i, v_j) = \frac{\mathbf{e}_i^\top \mathbf{e}_j}{\|\mathbf{e}_i\| \, \|\mathbf{e}_j\|}
\end{equation}
where $\mathbf{e}_i, \mathbf{e}_j\in\mathbf{H}^{(E)}$, the cosine similarity function $\operatorname{sim}(\cdot)$ can be replaced by dot product or other similarity calculations.

ii) \textbf{Popular Items $(v_i^{pop})$} appear more frequently and are often displayed with other content, increasing the likelihood of concurrent exposure~\cite{10.1145/3637528.3671824,10.1109/TKDE.2022.3218994}. Therefore, we ask ``\textit{What if the user was shown a very popular item?}'' To effectively model user preference under popular trends, we also construct the positively exposed item-time pair $(v^{pop}_i, t_i)$ with a counterfactual popular item. Specifically, we calculate the popularity score of items other than $v_i$ from the historical interaction data for all users $u\in\mathcal{U}$. Since we focus on modeling the exposure features of items and also to remain consistent with $v_i^{sim}$, we also retrieve the embedding of $v^{pop}_i$ from the exposure embedding matrix $\mathbf{H}^{(E)}$.
\begin{equation}
    \mathbf{e}^{pop}_i = \operatorname{Lookup}(\mathbf{H}^{(E)}, \operatorname{index}(v^{pop})) 
\end{equation}
where $v^{pop}\in\mathcal{V}$ is the selected popular item, $\operatorname{index}(\cdot)$ denotes the index function returning the id of $v^{pop}$. Specifically, we first count the occurrences of each item $v \in \mathcal{V}$ in $\mathcal{D}_{\text{train}}$, and then randomly sample one item $v^{pop}$ for each $v_i$ from the top-$K$ most popular items.
\begin{equation}
\begin{aligned}
    v^{pop} &\sim \operatorname{Unif}\!\left(\operatorname{TopK}\!\Big(\{\operatorname{count}(v)\mid v \in \mathcal{V}\}, K\Big)\right)\\
    \operatorname{count}(v) &= \sum_{(v_j, t_j) \in \mathcal{D}_{\text{train}}} \mathbf{1}\big(v_j = v \;\wedge\; t_j \in [t-\tau, t]\big)
\end{aligned}
\end{equation}
where $\operatorname{Unif}(\cdot)$ denotes that items are uniformly sampled from the rest of item set $\mathcal{V}\backslash v_i$, $K$ is a hyperparameter that controls the number of top popular items considered, $\mathcal{D}_{\text{train}}$ represents the training set of interaction data, $\tau$ controls the time window considered in the popularity counting function $\operatorname{count}(\cdot)$. 

iii) \textbf{Same Item, Different Time $(t_i^*)$} The same item is likely to be consistently exposed to a user within a short time window, due to mechanisms such as short-term re-ranking~\cite{10.1145/3477495.3532046}. Based on this findings, we ask ``\textit{What if the user was shown the same item, but at a slightly different moment?}''. To reasoning over the counterfactual exposure moment in this question, we applied a slight intervention $\Delta$ on the time embedding $\mathbf{t}_i$ to construct the item-time pair $(v_i, t_i^*)$ via $\mathbf{t}^*_i = \mathbf{t}_i + \Delta$. The $\Delta\in\mathbb{R}^{d}$ is a random vector of small magnitude. Here, we set $\Delta\in[-1 \times 10^{-4}, 1 \times 10^{-4}]$. 

Thus far, we have derived the embeddings for the three counterfactual item–time pairs $(\mathbf{e}_i^{sim}, \mathbf{t}_i)$, $(\mathbf{e}_i^{pop}, \mathbf{t}_i)$, and $(\mathbf{e}_i, \mathbf{t}_i^*)$. In this paper, these three counterfactual item–time pairs are regarded as positive samples for exposure distribution estimation in $f_\varphi$, but are treated as negative samples to augment interaction-based user preference inference in $g_\theta$. The following sections elaborate on how the three counterfactual pairs, together with the factual interaction $(v_i, t_i)$, are further leveraged in \name.

\subsection{Exposure Influence Interaction ($\mathbf{E}\rightarrow \mathbf{C}$)}
The \textbf{exposure bias} arises because users can only interact with items they have been exposed to, causing the observed interactions to reflect the exposure mechanism rather than pure user preference. Recall that the structural causal model in Figure \ref{fig:scm} shows that item exposure is a necessary prerequisite for user–item interaction. Motivated by this causal relationships, we incorporate exposure estimation into users’ historical interactions with temporal information and further adjust predictions using a time-aware inverse propensity scoring framework to mitigate exposure bias.

Given the embeddings of interacted items retrieved from the interaction embedding matrix $\mathbf{H}^{(C)}$, we arrange them according to their interaction timestamps $t_i$, resulting in the sequence $\{\mathbf{c}_0, \mathbf{c}_1, ..., \mathbf{c}_{M}\}$. We then combine this item sequence with the corresponding timestamp embeddings $\{\mathbf{t}_0, \mathbf{t}_1, ...\mathbf{t}_{M}\}$.

\begin{equation}
\mathbf{S}=\{\mathbf{c}_0, \mathbf{c}_1, ..., \mathbf{c}_{M}\}\oplus \{\mathbf{t}_0, \mathbf{t}_1, ...\mathbf{t}_{M}\}
\end{equation}
where $\oplus$ represents element-wise addition, $\mathbf{S}=\{\mathbf{s}_{m}\}\in\mathbb{R}^{M\times d}$ denotes the basic embedding of user preferences, derived solely from historical interactions while disregarding exposure information, and $M$ is the maximum length of user's historical interaction sequences, $m$ indexes the position of element in the sequence.

Before integration, to dynamically estimate the exposure distribution of the candidate items, we aggregate the item and embeddings for each factual or counterfactual item-time pairs, \ie $(v_{i}, t_{i})$, $(v^{sim}_{i}, t_{i})$, $(v^{pop}_{i}, t_{i})$, and $(v_{i}, t^*_{i})$. For example, for item-time pair $(v_{i}, t_{i})$, the hidden embedding $\mathbf{\hat{e}}_{i}\in\mathbb{R}^{d}$ is represented as below:
\begin{equation}
\mathbf{\hat{e}}_{i}=\mathbf{e}_{i}+\mathbf{t}_{i}
\end{equation}
where $\mathbf{e}_{i}$ and $\mathbf{t}_{i}$ are obtained from Sec. \ref{sec:item_enc} and Sec. \ref{sec:time_enc}, respectively. The symbol ``$+$'' represents the additive operation.
We then incorporate exposure information into the interaction sequence through the cross attention mechanism. Specifically, the exposure embedding $\mathbf{\hat{e}}_{i}$ serves as the query $\mathbf{q}$, while the embeddings of each interaction in the sequence $\mathbf{S}$ are employed as both key $\mathbf{k}$ and value $\mathbf{v}$. Formally, $\mathbf{q}^{(n)} = \mathbf{\hat{e}}_{i} \mathbf{W}_Q^{(n)} + \mathbf{b}_Q^{(n)}$, $\mathbf{k}^{(n)} = \mathbf{s}_{m} \mathbf{W}_K^{(n)} + \mathbf{b}_K^{(n)}$, and $\mathbf{v}^{(n)} = \mathbf{s}_{m} \mathbf{W}_V^{(n)} + \mathbf{b}_V^{(n)}$, 
where $\mathbf{W}\in\mathbb{R}^{\frac{d^2}{N}}$ and $\mathbf{b}\in\mathbb{R}^{\frac{d}{N}}$ denote the weighting matrix and bias, respectively. The $(n)$ represents the $n$-th head of cross-attention, $N$ denotes the total number of heads, and $\mathbf{s}_m$ denotes a subsequence of historical interactions for the same user. The cross attention at the $n$-th head is formulated as:
\begin{equation}
    \operatorname{Attention}(\mathbf{q}^{(n)}, \mathbf{k}^{(n)}, \mathbf{v}^{(n)})
    = \operatorname{softmax} \left( \frac{\mathbf{q}^{(n)} (\mathbf{k}^{(n)})^\top}{\sqrt{d}} \right) \mathbf{v}^{(n)}
\end{equation}

The output embedding of the cross-attention mechanism is aggregated via an average pooling operation to produce a scalar score. This score serves as the exposure propensity, representing the probability that item $v_i$ is exposed at time $t_i$.
\begin{equation}
    s_{v_i,t_i}=\pi^t(\mathbf{S}, \mathbf{\hat{e}}_i)= \operatorname{mean}(\mathbf{\hat{S}})
\label{eq:exp_propensity_s}
\end{equation}
\begin{equation}
\mathbf{\hat{S}}=\sum_n\operatorname{Concat}_m(\operatorname{Attention}(\mathbf{q}^{(n)}, \mathbf{k}^{(n)}, \mathbf{v}^{(n)}))
\label{eq:expo_aware_u}
\end{equation}
where $\pi^t(\cdot)$ denotes the function that estimates the time-aware exposure distribution, taking the interaction sequence $\mathbf{\hat{S}}$ and a specific exposure sample $\mathbf{\hat{e}}_i$ as inputs, and the subscript $m$ of $\text{Concat}(\cdot)$ denotes concatenating all the results from all the subsequences of historical interactions. The resulting time-aware propensity score $s$ of $\pi^t(\cdot)$ is used to address exposure bias (see Section \ref{sec:rec}).

To train the plug-in exposure estimation model $f_\varphi$ to accurately capture the exposure distribution and seamlessly integrate exposure information with user–item interactions, we construct a positive set $\mathcal{E}^+$ and a negative set $\mathcal{E}^-$. The positive set $\mathcal{E}^{+}$ includes items that are considered positively exposed along with their corresponding exposure times, consisting of the factual item-time pair $(v_i, t_i)$ observed in the training interactions and three artificially constructed counterfactual exposure samples for each pair: $(v_i^{sim}, t_i)$, $(v_i^{pop}, t_i)$, and $(v_i, t_i^*)$. The negative set $\mathcal{E}^{-}$ consists of randomly sampled item–time pairs $(v', t')$ drawn from the remaining data $\mathcal{D}\backslash\mathcal{E}^{+}$.
\begin{equation}
\label{eq:bce}
\mathcal{L}_{\text{EP}} = \frac{-1}{|\mathcal{E}|} 
\Bigg[
\sum_{(v,t) \in \mathcal{E}^+} \log \sigma\big(s_{v,t}\big) 
\\+ \sum_{(v',t')\in \mathcal{E}^-} \log \big(1 - \sigma\big(s_{v',t'}\big)\big)
\Bigg],
\end{equation}

\noindent
where $|\mathcal{E}|=|\mathcal{E}^{+}\cup\mathcal{E}^{-}|$ denotes the number of samples in the negative $\mathcal{E}^{-}=\{v', t'\}$ and positive sets $\mathcal{E}^+=\{(v_i,t_i),(v_i^{sim}, t_i),(v_i^{pop}, t_i),$\\$(v_i, t^*_i)\}$. The $s_{v,t}$ and $s_{v',t'}$ are the probability for positive sample $(v,t)$ and negative sample $v',t'$ to be exposed, respectively, computed with Eq. (\ref{eq:exp_propensity_s}). The $\sigma(\cdot)$ denotes the sigmoid function.

\subsection{Exposure Influence User Preference ($\mathbf{E}\rightarrow\mathbf{U}$)}
The \textbf{selection bias} arises because the recommender is trained solely on observed interactions in $\mathcal{D}_{\text{train}}$, which includes only items interacted with the user and can therefore skew the modeling of user preference. This bias often leads to overestimating the preference for popular or frequently exposed items while underrepresenting unobserved or less-exposed items. In this paper, we leverage the counterfactual item exposure data to provide additional information for learning user preferences, thereby mitigating the effects of selection bias and improving the generalization of the model, as illustrated by the path $\mathbf{E}\rightarrow\mathbf{U}$ in Fig. \ref{fig:scm}.

We regard our \name as a plug-in model designed to enhance exposure information for any SR models (denoted as $g_\theta$). To demonstrate the flexibility of \name, we illustrate how exposure effects can be modeled in two categories of recommendation systems: attention-based sequential models, exemplified by the simple attention layers~\cite{vaswani2017attention}, and generative models, exemplified by diffusion~\cite{diffurec}.

\subsubsection{Traditional Sequential Models} Following the common practice in SR, we apply a self-attention mechanism to the exposure-aware user preference $\mathbf{\hat{S}}_u$ (as computed in Eq. (\ref{eq:expo_aware_u})) to capture information between items in the interaction sequence. For modeling a user from their historical interaction sequence, the self-attention mechanism used in Eqs. (\ref{eq:attn_u}) can be replaced with any sequential model, such as an RNN or GRU modules.
\begin{equation}
    \mathbf{q}_u^{(n)}, \mathbf{k}_u^{(n)}, \mathbf{v}_u^{(n)} = \mathbf{\hat{s}}_{m} \mathbf{W}^{(n)} + \mathbf{b}^{(n)},\ 
    \mathbf{u} = \sum_n\operatorname{Attention}(\mathbf{q}_u^{(n)}, \mathbf{k}_u^{(n)}, \mathbf{v}_u^{(n)})
\label{eq:attn_u}
\end{equation}
where $\mathbf{q}_u^{(n)}$, $\mathbf{k}_u^{(n)}$, $\mathbf{v}_u^{(n)}$ denotes the query, key, value for calculating user embedding $\mathbf{u}$, $\mathbf{W}\in\mathbb{R}^{\frac{d^2}{N}}$, and $\mathbf{b}\in\mathbb{R}^{\frac{d}{N}}$.

Given the user preference embedding $\mathbf{u}$ and the interaction embedding $\mathbf{c}_i$ of candidate item $v_{i}$, the recommendation score $y_i$ can be predicted using their inner product, \ie, $y_{i} = \operatorname{InnerProduct}(\mathbf{u}, \mathbf{c}_i)$.

\subsubsection{Generative Model}
For the generative model represented by diffusion, we condition on the embedding of the exposure-aware user interaction sequence $\mathbf{\hat{S}}_u$ to generate the ideal item embedding $\mathbf{\hat{c}}_i$, which represents the user’s preference.

\begin{equation}
    \mathbf{\hat{c}}_i = Z_\Theta([\mathbf{z}_0, \mathbf{z}_1, ..., \mathbf{z}_M])
\end{equation}
\begin{equation}
    \mathbf{z}_i = \mathbf{\hat{s}}_{m} + \lambda_i \odot (\dot{\mathbf{c}}_i+\mathbf{d})
\end{equation}
where $Z_\Theta$ denotes the model to capture user preference from historical sequence, $\odot$ is the element-wise product, and $\mathbf{d}$ is the diffusion step embedding following the position encoding in ~\cite{vaswani2017attention}, $\lambda_i$ is sampled from a Gaussian distribution, $\mathbf{\hat{s}}_m$ is the exposure aware embedding for the interacted item at position $i$ in sequence $\mathbf{\hat{S}}$, and $\dot{\mathbf{c}}_i$ is the noised or reconstructed interaction embedding for candidate item $v_i$ during the forward diffusion or reverse process, respectively.

The recommendation score in diffusion-based models are calculated based on the similarity between the ideal item $\mathbf{\hat{v}}$ and items in the actual item set $\mathcal{V}$. For actual item $v_i$, there is, $y_i = \operatorname{Sim}(\mathbf{\hat{c}}_i, \mathbf{c}_i)
$.

\subsection{Recommendations}
\label{sec:rec}

Building on the modeling of exposure influence on interactions and user preferences, we propose a novel Time-aware Inverse Propensity Scoring (TIPS) method that incorporates temporal information into the final recommendation objective.

\subsubsection{Time-aware Inverse Propensity Scoring} For the factual item-time pair $(v_i,t_i)$ in the training interaction data, there is a time-aware inverse propensity score $\frac{1}{s_{v_i, t_i}}$. Therefore, the training objective of the SR model can be formulated as below.
 \begin{equation}
 \label{eq:bpr-tips}
     \mathcal{L}_{\text{BPR-TIPS}}=\frac{1}{N}\sum_{(v_i,t_i)\in\mathcal{C}^+, (v_j,t_j)\in\mathcal{C}^-}\frac{\mathbf{w}_{v_i,t_i}ln \sigma(y_{i}, y_{j})}{\sum_{(v_i,t_i)\in\mathcal{C}^+}s_{v_i,t_i}}
 \end{equation}

\noindent
where $y$ is calculated with inner product or similarity measurement, for sequential model-based or diffusion-based methods, respectively. The $s_{v_i,t_i}$ is the exposure propensity for $(v_i,t_i)\in\mathcal{C}^+$, as defined in Eq. (\ref{eq:exp_propensity_s}). Note that the factual item-time pair $(v_i, t_i)\in\mathcal{C}^+$ is regarded as the positive interaction sample, while the three counterfactual counterparts $(v_i^{sim}, t_i),(v_i^{pop}, t_i),(v_i, t^*_i)$ are regarded as the corresponding negative ones in the set $\mathcal{C}^-$.

In Eq. (\ref{eq:bpr-tips}), the  $\mathbf{w}_{v_i,t_i}$ incorporates time decaying with the exposure propensity score $s_{v_i, t_i}$ for item $v_i$ at time $t_i$, rather than applying direct adjustments. Specifically, $\mathbf{w}_{v_i,t_i}$ emphasizes recent interactions and further strengthens the correction for less frequently exposed items. Formally, the weight can be calculated as below.
\begin{equation}
    \mathbf{w}_{v_i,t_i} = \frac{\operatorname{exp}(-\mu(t_{i}-t_{i-1}))}{\operatorname{max}(s_{v_i,t_i}, \epsilon)}
\end{equation}
where $\mu>0$ is a decay hyperparameter that controls how fast the exponential term decreases with time, a smaller $\mu$ means a slower decay, so past interactions remain influential for a longer time. The $\epsilon$ is a small baseline value, ensuring stability and preventing division by very small exposure probabilities. 

\subsubsection{Model Optimization}
The proposed framework $f_\varphi\circ g_\theta$ is optimized through a combination of exposure distribution learning guided by $\mathcal{L}_{\text{EP}}$ in Eq. (\ref{eq:bce}) and recommendations adjusted with time-aware inverse propensity scoring with ($\mathcal{L}_{\text{BPR-TIPS}}$ in Eq. (\ref{eq:bpr-tips}). Formally, the final training objective is defined as follows.

\begin{equation}
    \mathcal{L} = \mathcal{L}_{BPR-TIPS}+\gamma\mathcal{L}_{EP}
\end{equation}
where $\gamma$ is a hyperparameter that controls the extent to which the \name framework focuses on item exposure. 

\section{Experiments}
To validate the effectiveness of \name for SR, we have conduct extensive experiments to answer the following research questions.
\begin{itemize}[leftmargin=*]
    \item \textbf{RQ1}: How does \name perform compared to state-of-the-art methods across different backbone recommendation models?
    \item \textbf{RQ2}: How do the components of \name improve sequential recommendation performance?
    \item \textbf{RQ3}: How do the different values of hyperparameters influence the recommendation performances?
    \item \textbf{RQ4}: In what ways does the proposed \name improve upon the traditional IPS?
\end{itemize}

\section{Experimental Setups}\label{sec:setups}
\subsubsection{Datasets} We evaluated our \name on four widely used recommendation datasets of varying sizes (Table \ref{tab:datasets_stats}). 
\begin{table}[ht]
  \caption{The statistics of datasets.}
  \vspace{-3mm}
  \centering
  \scriptsize
  \resizebox{\columnwidth}{!}{
  \begin{tabular}{c|cccc}
    \toprule
    Dataset &\#User&\#Item&\#Interaction&Time Period(mm/dd/yyyy)\\
    \midrule
    \bf ML-1M &5950 &3532 &574619 &04/25/2000-02/28/2003 (34.2 months)\\
    \bf ML-10M &66028 &10254 &4980475 &01/09/1995-01/05/2009 (168 months) \\
    \bf Music4All &14124 &99596 &5109577 &12/30/2013-03/26/2019 (62.8 months)\\
    \bf GoodReads &17660 &29972 &409600 &02/13/2001-11/02/2017 (200.3 months)\\
    \bottomrule
  \end{tabular}
}
\label{tab:datasets_stats}
\end{table}



Movielens (ML-1M/ML-10M)~\cite{movielens} includes rich metadata on movie items and user ratings over time, making it invaluable for evaluating and developing systems across different recommendation scenarios. 
Music4All~\cite{music4all} incorporates rich user–music interactions with precise timestamps, track release years, supporting research on both short-term dynamics and long-term temporal evolutions. GoodReads~\cite{goodreads1,goodreads2} collects user-book interactions in late 2017 from publicly available user shelves on goodreads.com.

\subsubsection{Evaluation Protocols} To evaluate recommendation performance, we adopt two commonly used metrics: HR@K evaluates whether at least one of the ground-truth interacted items appears within the top-$K$ recommended items. This score reflects the ability of the recommendation system to successfully retrieve user-interested items. NDCG@K measures the ranking quality of the top-$K$ items in the recommendation list by assigning higher importance to correctly ranked items at higher positions. 

\subsubsection{Baselines} We compare our proposed \name against both traditional sequential models~\cite{sasrec,tisasrec,gru4rec,vaswani2017attention} and generative models~\cite{cvae,tdiffrec,pdrec,diffurec,ddrm}, together with causal plug-in debiasing methods ~\cite{10.1145/3485447.3512092,11150713}.\\

\textit{Traditional Sequential Models}: \textbf{SASRec}~\cite{sasrec} uses self-attention to model long-range dependencies in  interactions.\textbf{TiSASRec}~\cite{tisasrec} incorporates time interval between interactions to better capture time-sensitive user preference. \textbf{GRU}~\cite{gru4rec} is a session-based recommendation model that leverages Gated Recurrent Units (GRUs) to capture sequential user behavior. \textbf{USR}~\cite{10.1145/3485447.3512092} debias selection and exposure bias with IPS based on GRUs. \textbf{Attention}~\cite{vaswani2017attention} applied the self-attention mechanism to historical interaction sequences and calculate the mean value as the final user preference. \textbf{LDPE}~\cite{11150713} debiases exposure bias with a LLM-enhanced attention architecture.

\textit{Generative Models}: \textbf{CVAE}~\cite{cvae} applies the conditional variational autoencoder to predict the next interacted item by conditioning on users’ historical interactions. \textbf{T-DiffRec}~\cite{tdiffrec} re-weights user interactions according to their timestamps to better capture temporal dynamics. \textbf{PDRec}~\cite{pdrec} models the dynamic preferences of users with a diffusion process of a time interval. \textbf{DiffuRec}~\cite{diffurec} utilizes a denoising diffusion probabilistic model to learn from intricate user–item interactions. \textbf{DDRM}~\cite{ddrm} leverages multi-step denoising process based on diffusion models to robustify embeddings.

\subsubsection{Implementation Details} We explore the learning rate within the range of \{5e-5, 1e-5, 5e-6\}, while searching batch sizes from \{8, 16, 32\}. The hyperparameters $\mu$, $\lambda$ from the range of \{0.05, 0.1, 0.2, 0.5, 1.0, 2.0\} and \{0.05, 0.1, 0.3, 0.5, 0.7, 1.0\}, respectively. All experiments are conducted on two NVIDIA RTX-A5000-24GB GPUs.

\subsection{Recommendation Performance (RQ1)}
\begin{table*}[t!]
\centering
\caption{Comparison of recommendation performance between \name and two categories of sequential models.}
\vspace{-3mm}
\resizebox{\textwidth}{!}{
\setlength{\tabcolsep}{2pt}
\begin{tabular}{lcccccccccccccccc}
\toprule
\multirow{2}{*}{\bf Model} &\multicolumn{4}{c}{\textbf{ML-1M}} &\multicolumn{4}{c}{\textbf{Music4All}} &\multicolumn{4}{c}{\textbf{GoodReads}} &\multicolumn{4}{c}{\textbf{ML-10M}} \\
\cmidrule(l){2-5}\cmidrule(l){6-9}\cmidrule(l){10-13}\cmidrule(l){14-17} 
 &HR@5 &HR@10 &NDCG@5 &NDCG@10 &HR@5 &HR@10 &NDCG@5 &NDCG@10 &HR@5 &HR@10 &NDCG@5 &NDCG@10 &HR@5 &HR@10 &NDCG@5 &NDCG@10 \\
 \midrule
 \multicolumn{17}{c}{\it Traditional Sequential Models}\\
 \midrule 
 SASRec &0.3351	&0.4501	&0.2249	&0.2836	&0.5581	&0.6469	&0.4119	&0.4473	&0.4814	&0.5571	&0.3622	&0.3877	&0.4755	&0.6319	&0.3164	&0.3704 \\
 TiSASRec &0.3525 &0.4819 &0.2387	&0.2968	&0.5808	&0.6924	&0.4429	&\underline{0.4772}	&\underline{0.4981}	&\underline{0.6017}	&0.3812	&0.4113	&0.4987	&0.6721	&0.3375	&0.3924\\

GRU (Base)
&0.3467 &0.4731 &0.2482 &0.2895
&0.5460 &0.6253 &0.3983 &0.4336
&0.4607 &0.5372 &0.3501 &0.3732
&0.4593 &0.6027 &0.3061 &0.3538 \\
+USR 
&0.3728 &0.5123 &0.2658 &0.3119
&0.5762 &0.6501 &0.4173 &0.4581
&0.4763 &0.5567 &0.3627 &0.3862
&0.4941 &0.6502 &0.3260 &0.3807 \\
\rowcolor{myblue}
+\name (Ours)
&\underline{0.3902} &\underline{0.5317} &0.2808 &\underline{0.3269}
&\underline{0.5932} &0.6716 &0.4322 &0.4728
&0.4970 &0.5842 &\underline{0.3940} &0.4129
&0.5377 &0.7171 &0.3605 &0.4167 \\
\rowcolor{myblue} Improv. &\up{4.35\%}&\up{5.86\%}&\up{3.26\%}&\up{3.74\%}&\up{4.72\%}&\up{4.63\%}&\up{3.39\%}&\up{3.92\%}&\up{3.63\%}&\up{4.70\%}&\up{4.39\%}&\up{3.97\%}&\up{7.84\%}&\up{11.44\%}&\up{5.44\%}&\up{6.29\%}\\

 Attn. (Base)&0.3666	&0.4951	&0.2662	&0.3075	&0.5404	&0.6592	&0.4069	&0.4453	&0.4733	&0.5738	&0.3657	&0.3982	&0.6123	&0.7503	&0.4492	&0.4921\\
 +LDPE&0.3689&0.5099&\underline{0.2882}&0.3236&0.5547&\underline{0.6903}&\underline{0.4458}&0.4691&0.4772&0.5983&0.3782&\underline{0.4230}&\underline{0.6271}&\underline{0.7652}&\underline{0.4609}&\underline{0.5117}\\
 \rowcolor{myblue}+\name(Ours)  &\bf 0.4086	&\bf 0.5558	&\bf 0.2944	&\bf 0.3418	&\bf 0.6155	&\bf 0.7295	&\bf 0.4956	&\bf 0.5325	&\bf 0.5077	&\bf 0.6054	&\bf 0.4057	&\bf 0.4273	&\bf 0.6491	&\bf 0.7750	&\bf 0.4806	&\bf 0.5355\\ 
   \rowcolor{myblue} Improv. &\up{4.20\%} & \up{6.07\%} &\up{2.82\%} & \up{3.43\%} & \up{7.51\%} & \up{7.03\%} & \up{8.87\%} & \up{8.72\%} & \up{3.44\%} & \up{3.16\%} & \up{4.00\%} & \up{2.91\%} & \up{3.68\%} & \up{2.47\%} & \up{3.14\%} & \up{4.34\%}\\

 \midrule
\multicolumn{17}{c}{\it Generative Models}\\
 \midrule 
  DDRM &0.3999	&0.5501	&0.2545	&0.3218	&0.5984	&0.7256	&0.4639	&0.5009	&0.5027	&0.6326	&0.3652	&0.4027	&0.5324	&0.7119	&0.3709	&0.4246\\
 T-DiffRec &0.3336	&0.4624	&0.2206	&0.2717	&0.5339	&0.6427	&0.4179	&0.4531	&0.4873	&0.5787	&0.3628	&0.4024	&0.4444	&0.6077	&0.3091	&0.3619\\
 PDRec &0.3374	&0.486	&0.2259	&0.2805	&0.5453	&0.6517	&0.4207	&0.4598	&0.4879	&0.5877	&0.3669	&0.4127	&0.4712	&0.6278	&0.3127	&0.3875\\
 CVAE &0.3391	&0.4503	&0.2497	&0.2898	&0.4957	&0.5967	&0.3652	&0.3999	&0.4468	&0.5435	&0.3354	&0.3735	&0.5941	&0.7242	&0.4422	&0.4685\\
\rowcolor{myblue} +\name (Ours)  &0.3785	&0.4954	&0.2893	&0.3366	&0.5629	&0.6623	&0.4221	&0.4457	&0.4844	&0.5799	&0.3789	&0.4123	&0.6584	&0.7786	&0.4995	&0.5088\\
\rowcolor{myblue} Improv. &\up{3.94\%} &\up{4.51\%} &\up{3.96\%} &\up{4.68\%} & \up{6.72\%} &\up{6.56\%} &\up{5.69\%} &\up{4.58\%} & \up{3.76\%} &\up{3.64\%} &\up{4.35\%} & \up{3.88\%} & \up{6.43\%} &\up{5.44\%} & \up{5.73\%} & \up{4.03\%}\\
DiffuRec &\underline{0.4801}	&\underline{0.6311}	&\underline{0.3209}	&\underline{0.3665}	&\underline{0.6304}	&\underline{0.7341}	&\underline{0.4856}	&\underline{0.5294}	&\underline{0.5302}	&\underline{0.6673}	&\underline{0.3822}	&\underline{0.4267}	&\underline{0.7162}	&\underline{0.8327}	&\underline{0.5189}	&\underline{0.5927}\\
 \rowcolor{myblue} +\name (Ours)  &\bf 0.5173	&\bf 0.6712	&\bf 0.3505	&\bf 0.4072	&\bf 0.6844	&\bf 0.7964	&\bf 0.5391	&\bf 0.5817	&\bf 0.5703	&\bf 0.6885	&\bf 0.4169	&\bf 0.4653	&\bf 0.7591	&\bf 0.8794	&\bf 0.5778	&\bf 0.6171\\
  \rowcolor{myblue}Improv. &\up{3.72\%} &\up{4.01\%} &\up{2.96\%} &\up{4.07\%} &\up{5.40\%} &\up{6.23\%} &\up{5.35\%} &\up{5.23\%} &\up{4.01\%} &\up{2.12\%} &\up{3.47\%} &\up{3.86\%} &\up{4.29\%} &\up{4.67\%} &\up{5.89\%} &\up{2.44\%}\\
\bottomrule 
\end{tabular}}
\label{tab:baselines}
\end{table*}

Table \ref{tab:baselines} compares \name with 7 baseline models, incorporating traditional sequential models ~\cite{vaswani2017attention,sasrec,tisasrec} and generative models ~\cite{cvae,tdiffrec,pdrec,diffurec,ddrm}. We applied \name to three backbone models: a traditional sequential recommender based on the attention mechanism~\cite{vaswani2017attention}, and two generative recommenders, CVAE~\cite{cvae} and DiffuRec~\cite{diffurec}.

\textbf{Our proposed \name framework improves the recommendation performance on both traditional sequential models and generative models across different datasets.} As shown in Table \ref{tab:baselines}, \name on three backbone models achieves competing performance on the four datasets with different scales and varied scenarios. For the traditional sequential model based on the attention mechanism~\cite{vaswani2017attention}, incorporating TIPS (Attention+TIPS) achieves an average improvement of around 6\% in HR@10 and 5\% in NDCG@10. Generative models such as CVAE~\cite{cvae} and DiffuRec~\cite{diffurec} also benefit from TIPS, with relative gains of approximately 5\% in HR@10. 
    
\textbf{The performance improvement of TIPS for SR models is greater on large-scale datasets than on small-scale datasets} Notably, \name performs particularly well on improving the performance on Music4All and ML-10M in Table \ref{tab:baselines}, indicating the superior improvements of \name on diverse and large-scale recommendation scenarios. For example, \name achieves an improvement of up to 8.87\% in HR@10, along with a comparably strong gain of 8.72\% in NDCG@10. We contribute this improvements on the active users and rich interactions in the MusicAll dataset. The user engagement on Music4All, with 5.76 monthly interactions per user, is significantly higher than the average of 1.14 observed across the other three datasets. By leveraging a greater number of user interactions, Inverse Propensity Scoring (IPS) can achieve more precise correction of selection bias, which in turn helps alleviate exposure bias. This process facilitates a more comprehensive learning of true user preferences, leading to a significant improvement in the effectiveness of the recommendations.

\textbf{The incorporation of temporal information proven to be effective on SRs.} In Table \ref{tab:baselines}, models incorporating temporal information consistently outperforms their counterparts  without modeling the time information. For example, TiSASRec achieves average improvements of 5.84\% and 6.39\% in HR@10 and NDCG@10, respectively. This is aligned with our motivations that fully leveraging the temporal information to enhance the adjustment of inverse propensity scoring. Similarly, our \name benefits from modeling temporal dynamics, with \name achieving even larger gains. Specifically, the attention mechanism with \name achieves comparable performance (over 40\% in HR@10) in complex recommendation scenarios (\ie Music4All, GoodReads, and ML-10M) to generative models, and significantly outperforms the traditional sequential model with temporal information (\ie Ti-SASRec). 

\subsection{Ablation Study (RQ2)}
\begin{table}[t!]
\centering
\caption{Ablation studies of \name on the ML-1M.}
\vspace{-3mm}
\resizebox{0.8\columnwidth}{!}{
\setlength{\tabcolsep}{2pt}
\begin{tabular}{clcccc}
\toprule
\multirow{2}{*}{\bf Backbone}&\multirow{2}{*}{\bf Variant} &\multicolumn{4}{c}{\textbf{ML-1M}} \\
\cmidrule(l){3-6}
&&\multicolumn{2}{c}{HR@10} &\multicolumn{2}{c}{NDCG@10} \\
 \midrule
 \multirow{4}{*}{Attention}& \name(Ours) &0.5558&&0.3418& \\ 
 &\name$\neg \text{time}$ &0.5550&\downbad{0.08\%}&0.3472&\up{0.54\%} \\
 &\name$\neg \text{IPS}$ &0.5356&\downbad{2.02\%}&0.3238&\downbad{1.80\%}\\
 &\name$\neg \text{EP\&time}$ &0.5311&\downbad{2.47\%}&0.2874&\downbad{5.44\%}\\
 \midrule
\multirow{4}{*}{CVAE}& \name(Ours)   &0.4954&&0.3366& \\ 
 &\name$\neg \text{time}$ &0.4911&\downbad{0.43\%}&0.3311&\downbad{0.55\%}\\
 &\name$\neg \text{IPS}$ &0.4855&\downbad{0.99\%}&0.3219&\downbad{1.47\%}\\
 &\name$\neg \text{EP\&time}$ &0.4415&\downbad{5.39\%}&0.2757&\downbad{6.09\%}\\
 \midrule
 \multirow{4}{*}{Diffusion}& \name(Ours) &0.6712&&0.4072& \\ 
 &\name$\neg \text{time}$ &0.6581&\downbad{1.31\%}&0.3875&\downbad{1.97\%}\\
 &\name$\neg \text{IPS}$ &0.6537&\downbad{1.75\%}&0.3889&\downbad{1.83\%}\\
 &\name$\neg \text{EP\&time}$ &0.6217&\downbad{4.95\%}&0.3578&\downbad{4.94\%}\\
\bottomrule
\end{tabular}}
\label{tab:ablation_short}
\vspace{-5mm}
\end{table}

We conduct ablation studies with the following variants of \name: 
1) \textbf{\name{}$\neg\text{time}$} removes temporal information from both exposure estimation and inverse propensity adjustment, while preserving the overall framework with an end-to-end training strategy. 2) \textbf{\name{}$\neg\text{IPS}$} discards the inverse propensity adjustment and relies solely on the BPR loss for recommendation. The exposure estimation module is pretrained with temporal information using the BCE loss, and the whole framework is optimized in an end-to-end manner. 3) \textbf{\name{}$\neg\text{EP\&time}$} removes both the exposure estimation module and temporal information, using a pretrained propensity score, and is essentially a standard IPS-based method.

Table \ref{tab:ablation_short} shows these variants’ performance on ML-10M. It can be observed that all three variants perform worse than the original \name. This highlights both the necessity and the comprehensiveness of the \name design. The variant \name{}$\neg\text{EP\&time}$, which removes both the exposure estimation module and temporal information while relying only on IPS adjustments, suffers the largest performance drop
This demonstrates that both exposure modeling and temporal information are critical for recommendation performance in datasets of different scales, especially for achieving accurate ranking. In comparison, the variants \name{}$\neg\text{time}$ and \name{}$\neg\text{IPS}$ contribute less to overall performance degradation. This contrast underscores the importance and necessity of integrating temporal information with item counterfactual exposure estimation, further validating the motivation behind our \name. However, generative recommenders are more resilient to the removal of temporal information compared to traditional sequential models. Specifically, the variant \name{}$\neg\text{time}$ shows average declines of only 0.59\% and 1.03\% in HR@10 for the diffusion model~\cite{diffurec} and CVAE~\cite{cvae}, respectively, whereas the attention model~\cite{vaswani2017attention} suffers a larger decline of 2.68\%. This may be because generative models capture signals akin to temporal information, which provide extra cues for inference.

\subsection{Hyperparameter Study (RQ3)}

\begin{figure*}[htbp]
    \centering
    \includegraphics[width=\textwidth]{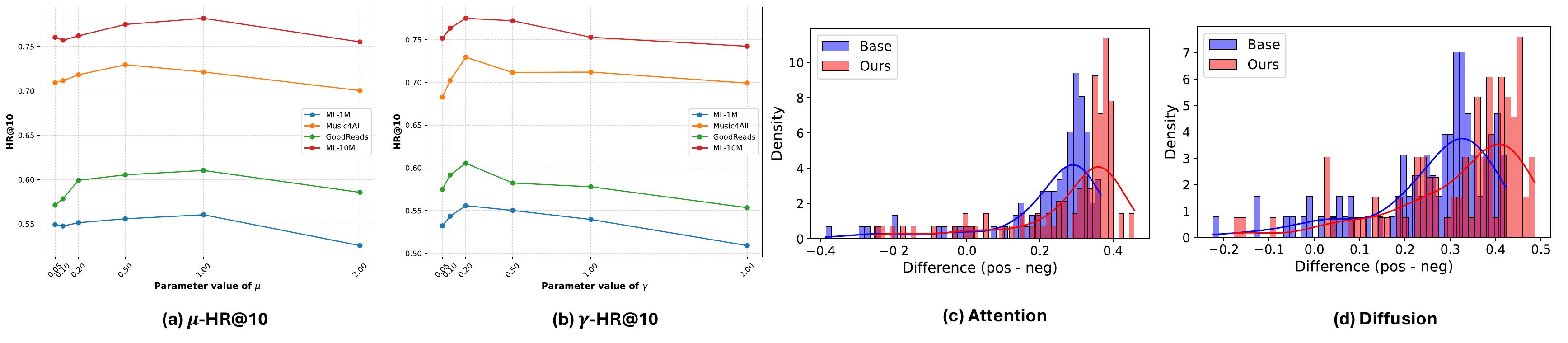}
    \vspace{-8mm}
    \caption{(a)(b) Results for different values of hyperparameters $\mu$ and $\gamma$, and (c)(d) distribution of the difference between the average propensity score of positive items and that of negative items, using \name (Ours) and traditional IPS (Base).}
    \label{fig:rq34}
\end{figure*}


To further evaluate the effectiveness of \name, we compute the recommendation performance of \name on the attention backbone with respect to HR@10 and NDCG@10 with the hyperparameters $\mu$ from \{0.05, 0.10, 0.20, 0.50, 1.00, 2.00\} and $\gamma$ from \{0.05, 0.10, 0.20, 0.50, 1.00, 2.00\}, respectively. We conduct experiments on the ML-1M with attantion as backbone and present the performance in Fig \ref{fig:rq34}(a)(b). 

The \textbf{hyperparameter $\mu$} is a decaying ratio that controls how quickly historical user interactions lose influence as they become outdated. A smaller $\mu$ produces a slower decay, allowing long-term behaviors to continue contributing to user modeling, wheareas a larger $\mu$ emphasizes only the most recent actions. As shown in Fig. \ref{fig:rq34} (a), performance rises as $\mu$ increases from 0.05 to about 0.20–0.50, indicating that moderately emphasizing recency improves recommendation quality, but further increases (when $\mu$ $\geq$ 1.0) reduce HR@10. This suggests that overly aggressive decay lost useful long-term signals. We set $\mu=0.5$ in our experiments. The \textbf{hyperparameter $\gamma$} controls the contribution of the exposure estimation loss $\mathcal{L}_{EP}$ to the overall training objective. The optimizer with a smaller $\gamma$ focuses primarily on fitting observed historical interactions, with minimal correction for exposure bias. Figures \ref{fig:rq34} (b) show that HR@10 improves as $\gamma$ increases to around 0.10, indicating that moderate incorporation of exposure information enhances recommendation predictions. In this paper, we set $\gamma$ to 0.3.

\subsection{Comparative Analysis (RQ4)}
In this section, we conduct a detailed analysis of the performance of \name compared to the traditional IPS method on the ML-1M dataset. We evaluate \name and IPS on three backbone recommenders: Attention~\cite{vaswani2017attention} for traditional sequential models, and diffusion~\cite{diffurec} for generative models. Specifically, we randomly sample 100 users and record the propensity scores computed by our \name (via $\pi^t(\cdot)$ in Eq. (\ref{eq:expo_aware_u})) and by the traditional IPS method for 100 items per user, including one positive item from the test set and 99 negative items randomly drawn from the remaining item pool. Figure \ref{fig:rq34}(c)(d) shows the distribution of the differences between the propensity scores of positive items and the average propensity scores of negative items. 

The propensity score inherently estimates the probability of an item being interacted with, while also incorporating the exposure information into account, \eg encoded in $\mathbf{\hat{e}}_i$ Eq. (\ref{eq:expo_aware_u}). The three backbone models then leverage these propensity scores to adjust for bias. \textbf{As shown in Fig. \ref{fig:rq34}(c)(d), \name consistently provides more discriminative propensity scores between positive and negative items, relative to the traditional IPS method.} More discriminative propensity scores for positive items is beneficial for mitigating exposure bias. Higher discriminability of positive-item propensity scores helps mitigate exposure bias: a larger difference indicates the model better identifies observed positives, allowing proper reweighting of less-exposed items. This also benefits selection bias, as it demonstrates the model’s ability to recognize exposed but not interacted items and assign them appropriate scores in the final objective function through inverse propensity weighting.
\section{Related Works}
\subsection{Sequential Recommendations}
Sequential recommendations (SRs) aim at predicting users' next interaction based on their historical interactions with items. Existing sequential recommenders can be broadly categorized into traditional sequential models and generative models. Traditional sequential models focus on learning patterns from user interaction histories by modeling the order and temporal dependencies among items. Recent techniques adopt deep neural architectures such as recurrent neural networks~\cite{sasrec,gru4rec}, convolutional networks~\cite{jiang2023adamct}, and attention-based transformers~\cite{sun2019bert4rec} to capture the sequential information of user preferences from interaction sequences. Differently, generative models treat SR as a generation problem, where the model directly generates the next item conditioned on user history, Leveraging frameworks such as variational autoencoders~\cite{10.1145/3511808.3557268,10.1145/3442381.3449873}, diffusion model~\cite{ddrm,tdiffrec}, and large language model~\cite{10.1145/3589335.3648307}. For example, Li \etal ~\cite{diffurec} leverages noise and reverse processes to better capture uncertainty in user preferences. Generative models are more flexible than traditional sequential models in adapting to shifts in user preferences, yet both suffer from the lack of exposure logs, leading to biased estimates of true preferences.

\subsection{Causality for Sequential Recommendations}
Causality-based methods~\cite{becausal,10.1016/j.neucom.2023.01.089} demonstrate the effectiveness of explicitly modeling the causal structure underlying observed interactions, rather than relying solely on model parameters to capture correlations. To mitigate biases arising from latent confounders in the causal structure, many of these methods adopt inverse propensity scoring (IPS)~\cite{10.1145/3488560.3498375,10.1145/3437963.3441799,10.1145/3366423.3380255,10.1016/j.eswa.2022.118932,ma2023selection} to reweight interactions during training. Wang \etal ~\cite{10.1145/3485447.3512092} utilize the inverse propensity scores to reweight the interactions in training data to debias SRs from latent confounders such as selection bias. Yu \etal ~\cite{11150713} integrates time-aware debiased propensity scores from both the item and user sides with the semantics understanding capability of LLMs to improve SRs. Similar to our work, CaseRec~\cite{10.1145/3726302.3730005} also focuses on inadequate system exposure data and leverages counterfactual augmentation to explore latent user interests. Though some existing causality-based methods incorporate temporal information to address bias, our approach differs by providing explicit guidance for exposure estimation through time-aware counterfactual samples.

\section{Conclusion}
In this paper, we propose Time-aware Inverse Propensity Scoring (\name), a causality-driven framework for sequential recommendation that mitigates biases from missing exposure data, notably exposure and selection bias. \name constructs counterfactual item–time pairs to estimate exposure distributions with temporal dynamics, reweighting user interactions via time-sensitive propensity scores. We evaluate \name on four public datasets against two categories of state-of-the-art models: (i) traditional sequential recommenders (RNN/attention-based) and (ii) generative models (VAEs, diffusion). Across three backbone models, \name consistently improves performance. As a versatile plug-in, it enables debiasing for existing sequential recommenders in both research and practical settings where exposure data are unavailable.

\clearpage










\begin{thebibliography}{43}


\ifx \showCODEN    \undefined \def \showCODEN     #1{\unskip}     \fi
\ifx \showISBNx    \undefined \def \showISBNx     #1{\unskip}     \fi
\ifx \showISBNxiii \undefined \def \showISBNxiii  #1{\unskip}     \fi
\ifx \showISSN     \undefined \def \showISSN      #1{\unskip}     \fi
\ifx \showLCCN     \undefined \def \showLCCN      #1{\unskip}     \fi
\ifx \shownote     \undefined \def \shownote      #1{#1}          \fi
\ifx \showarticletitle \undefined \def \showarticletitle #1{#1}   \fi
\ifx \showURL      \undefined \def \showURL       {\relax}        \fi
\providecommand\bibfield[2]{#2}
\providecommand\bibinfo[2]{#2}
\providecommand\natexlab[1]{#1}
\providecommand\showeprint[2][]{arXiv:#2}

\bibitem[Cai et~al\mbox{.}(2024)]%
        {10.1145/3637528.3671824}
\bibfield{author}{\bibinfo{person}{Miaomiao Cai}, \bibinfo{person}{Lei Chen}, \bibinfo{person}{Yifan Wang}, \bibinfo{person}{Haoyue Bai}, \bibinfo{person}{Peijie Sun}, \bibinfo{person}{Le Wu}, \bibinfo{person}{Min Zhang}, {and} \bibinfo{person}{Meng Wang}.} \bibinfo{year}{2024}\natexlab{}.
\newblock \showarticletitle{Popularity-Aware Alignment and Contrast for Mitigating Popularity Bias}. In \bibinfo{booktitle}{\emph{Proceedings of the 30th ACM SIGKDD Conference on Knowledge Discovery and Data Mining}} (Barcelona, Spain) \emph{(\bibinfo{series}{KDD '24})}. \bibinfo{publisher}{Association for Computing Machinery}, \bibinfo{address}{New York, NY, USA}, \bibinfo{pages}{187–198}.
\newblock
\showISBNx{9798400704901}
\href{https://doi.org/10.1145/3637528.3671824}{doi:\nolinkurl{10.1145/3637528.3671824}}


\bibitem[Fan et~al\mbox{.}(2022)]%
        {fan2022comprehensive}
\bibfield{author}{\bibinfo{person}{Wenqi Fan}, \bibinfo{person}{Xiangyu Zhao}, \bibinfo{person}{Xiao Chen}, \bibinfo{person}{Jingran Su}, \bibinfo{person}{Jingtong Gao}, \bibinfo{person}{Lin Wang}, \bibinfo{person}{Qidong Liu}, \bibinfo{person}{Yiqi Wang}, \bibinfo{person}{Han Xu}, \bibinfo{person}{Lei Chen}, {et~al\mbox{.}}} \bibinfo{year}{2022}\natexlab{}.
\newblock \showarticletitle{A comprehensive survey on trustworthy recommender systems}.
\newblock \bibinfo{journal}{\emph{arXiv preprint arXiv:2209.10117}} (\bibinfo{year}{2022}).
\newblock


\bibitem[Gao et~al\mbox{.}(2023)]%
        {10.1145/3568022}
\bibfield{author}{\bibinfo{person}{Chen Gao}, \bibinfo{person}{Yu Zheng}, \bibinfo{person}{Nian Li}, \bibinfo{person}{Yinfeng Li}, \bibinfo{person}{Yingrong Qin}, \bibinfo{person}{Jinghua Piao}, \bibinfo{person}{Yuhan Quan}, \bibinfo{person}{Jianxin Chang}, \bibinfo{person}{Depeng Jin}, \bibinfo{person}{Xiangnan He}, {and} \bibinfo{person}{Yong Li}.} \bibinfo{year}{2023}\natexlab{}.
\newblock \showarticletitle{A Survey of Graph Neural Networks for Recommender Systems: Challenges, Methods, and Directions}.
\newblock \bibinfo{journal}{\emph{ACM Trans. Recomm. Syst.}} \bibinfo{volume}{1}, \bibinfo{number}{1}, Article \bibinfo{articleno}{3} (\bibinfo{date}{March} \bibinfo{year}{2023}), \bibinfo{numpages}{51}~pages.
\newblock
\href{https://doi.org/10.1145/3568022}{doi:\nolinkurl{10.1145/3568022}}


\bibitem[Harper and Konstan(2015)]%
        {movielens}
\bibfield{author}{\bibinfo{person}{F.~Maxwell Harper} {and} \bibinfo{person}{Joseph~A. Konstan}.} \bibinfo{year}{2015}\natexlab{}.
\newblock \showarticletitle{The MovieLens Datasets: History and Context}.
\newblock \bibinfo{journal}{\emph{ACM Trans. Interact. Intell. Syst.}} \bibinfo{volume}{5}, \bibinfo{number}{4}, Article \bibinfo{articleno}{19} (\bibinfo{date}{Dec.} \bibinfo{year}{2015}), \bibinfo{numpages}{19}~pages.
\newblock
\showISSN{2160-6455}
\href{https://doi.org/10.1145/2827872}{doi:\nolinkurl{10.1145/2827872}}


\bibitem[Harvey et~al\mbox{.}(2021)]%
        {cvae}
\bibfield{author}{\bibinfo{person}{William Harvey}, \bibinfo{person}{Saeid Naderiparizi}, {and} \bibinfo{person}{Frank Wood}.} \bibinfo{year}{2021}\natexlab{}.
\newblock \showarticletitle{Conditional image generation by conditioning variational auto-encoders}.
\newblock \bibinfo{journal}{\emph{arXiv preprint arXiv:2102.12037}} (\bibinfo{year}{2021}).
\newblock


\bibitem[Hidasi et~al\mbox{.}(2015)]%
        {gru4rec}
\bibfield{author}{\bibinfo{person}{Bal{\'a}zs Hidasi}, \bibinfo{person}{Alexandros Karatzoglou}, \bibinfo{person}{Linas Baltrunas}, {and} \bibinfo{person}{Domonkos Tikk}.} \bibinfo{year}{2015}\natexlab{}.
\newblock \showarticletitle{Session-based recommendations with recurrent neural networks}.
\newblock \bibinfo{journal}{\emph{arXiv preprint arXiv:1511.06939}} (\bibinfo{year}{2015}).
\newblock


\bibitem[Hu et~al\mbox{.}(2024)]%
        {10.1145/3589335.3648307}
\bibfield{author}{\bibinfo{person}{Jun Hu}, \bibinfo{person}{Wenwen Xia}, \bibinfo{person}{Xiaolu Zhang}, \bibinfo{person}{Chilin Fu}, \bibinfo{person}{Weichang Wu}, \bibinfo{person}{Zhaoxin Huan}, \bibinfo{person}{Ang Li}, \bibinfo{person}{Zuoli Tang}, {and} \bibinfo{person}{Jun Zhou}.} \bibinfo{year}{2024}\natexlab{}.
\newblock \showarticletitle{Enhancing Sequential Recommendation via LLM-based Semantic Embedding Learning}. In \bibinfo{booktitle}{\emph{Companion Proceedings of the ACM Web Conference 2024}} (Singapore, Singapore) \emph{(\bibinfo{series}{WWW '24})}. \bibinfo{publisher}{Association for Computing Machinery}, \bibinfo{address}{New York, NY, USA}, \bibinfo{pages}{103–111}.
\newblock
\showISBNx{9798400701726}
\href{https://doi.org/10.1145/3589335.3648307}{doi:\nolinkurl{10.1145/3589335.3648307}}


\bibitem[Huang et~al\mbox{.}(2022)]%
        {10.1145/3488560.3498375}
\bibfield{author}{\bibinfo{person}{Jin Huang}, \bibinfo{person}{Harrie Oosterhuis}, {and} \bibinfo{person}{Maarten de Rijke}.} \bibinfo{year}{2022}\natexlab{}.
\newblock \showarticletitle{It Is Different When Items Are Older: Debiasing Recommendations When Selection Bias and User Preferences Are Dynamic}. In \bibinfo{booktitle}{\emph{Proceedings of the Fifteenth ACM International Conference on Web Search and Data Mining}} (Virtual Event, AZ, USA) \emph{(\bibinfo{series}{WSDM '22})}. \bibinfo{publisher}{Association for Computing Machinery}, \bibinfo{address}{New York, NY, USA}, \bibinfo{pages}{381–389}.
\newblock
\showISBNx{9781450391320}
\href{https://doi.org/10.1145/3488560.3498375}{doi:\nolinkurl{10.1145/3488560.3498375}}


\bibitem[Jiang et~al\mbox{.}(2023)]%
        {jiang2023adamct}
\bibfield{author}{\bibinfo{person}{Juyong Jiang}, \bibinfo{person}{Peiyan Zhang}, \bibinfo{person}{Yingtao Luo}, \bibinfo{person}{Chaozhuo Li}, \bibinfo{person}{Jae~Boum Kim}, \bibinfo{person}{Kai Zhang}, \bibinfo{person}{Senzhang Wang}, \bibinfo{person}{Xing Xie}, {and} \bibinfo{person}{Sunghun Kim}.} \bibinfo{year}{2023}\natexlab{}.
\newblock \showarticletitle{AdaMCT: adaptive mixture of CNN-transformer for sequential recommendation}. In \bibinfo{booktitle}{\emph{Proceedings of the 32nd ACM international conference on information and knowledge management}}. \bibinfo{pages}{976--986}.
\newblock


\bibitem[Kang and McAuley(2018)]%
        {sasrec}
\bibfield{author}{\bibinfo{person}{Wang-Cheng Kang} {and} \bibinfo{person}{Julian McAuley}.} \bibinfo{year}{2018}\natexlab{}.
\newblock \showarticletitle{Self-attentive sequential recommendation}. In \bibinfo{booktitle}{\emph{2018 IEEE international conference on data mining (ICDM)}}. IEEE, \bibinfo{pages}{197--206}.
\newblock


\bibitem[Kawar et~al\mbox{.}(2022)]%
        {ddrm}
\bibfield{author}{\bibinfo{person}{Bahjat Kawar}, \bibinfo{person}{Michael Elad}, \bibinfo{person}{Stefano Ermon}, {and} \bibinfo{person}{Jiaming Song}.} \bibinfo{year}{2022}\natexlab{}.
\newblock \showarticletitle{Denoising diffusion restoration models}.
\newblock \bibinfo{journal}{\emph{Advances in neural information processing systems}}  \bibinfo{volume}{35} (\bibinfo{year}{2022}), \bibinfo{pages}{23593--23606}.
\newblock


\bibitem[Li et~al\mbox{.}(2020)]%
        {tisasrec}
\bibfield{author}{\bibinfo{person}{Jiacheng Li}, \bibinfo{person}{Yujie Wang}, {and} \bibinfo{person}{Julian McAuley}.} \bibinfo{year}{2020}\natexlab{}.
\newblock \showarticletitle{Time Interval Aware Self-Attention for Sequential Recommendation}. In \bibinfo{booktitle}{\emph{Proceedings of the 13th International Conference on Web Search and Data Mining}} (Houston, TX, USA) \emph{(\bibinfo{series}{WSDM '20})}. \bibinfo{publisher}{Association for Computing Machinery}, \bibinfo{address}{New York, NY, USA}, \bibinfo{pages}{322–330}.
\newblock
\showISBNx{9781450368223}
\href{https://doi.org/10.1145/3336191.3371786}{doi:\nolinkurl{10.1145/3336191.3371786}}


\bibitem[Li et~al\mbox{.}(2023b)]%
        {becausal}
\bibfield{author}{\bibinfo{person}{Qian Li}, \bibinfo{person}{Xiangmeng Wang}, \bibinfo{person}{Zhichao Wang}, {and} \bibinfo{person}{Guandong Xu}.} \bibinfo{year}{2023}\natexlab{b}.
\newblock \showarticletitle{Be Causal: De-Biasing Social Network Confounding in Recommendation}.
\newblock \bibinfo{journal}{\emph{ACM Trans. Knowl. Discov. Data}} \bibinfo{volume}{17}, \bibinfo{number}{1}, Article \bibinfo{articleno}{14} (\bibinfo{date}{Feb.} \bibinfo{year}{2023}), \bibinfo{numpages}{23}~pages.
\newblock
\showISSN{1556-4681}
\href{https://doi.org/10.1145/3533725}{doi:\nolinkurl{10.1145/3533725}}


\bibitem[Li et~al\mbox{.}(2023a)]%
        {diffurec}
\bibfield{author}{\bibinfo{person}{Zihao Li}, \bibinfo{person}{Aixin Sun}, {and} \bibinfo{person}{Chenliang Li}.} \bibinfo{year}{2023}\natexlab{a}.
\newblock \showarticletitle{DiffuRec: A Diffusion Model for Sequential Recommendation}.
\newblock \bibinfo{journal}{\emph{ACM Trans. Inf. Syst.}} \bibinfo{volume}{42}, \bibinfo{number}{3}, Article \bibinfo{articleno}{66} (\bibinfo{date}{Dec.} \bibinfo{year}{2023}), \bibinfo{numpages}{28}~pages.
\newblock
\showISSN{1046-8188}
\href{https://doi.org/10.1145/3631116}{doi:\nolinkurl{10.1145/3631116}}


\bibitem[Liao et~al\mbox{.}(2024)]%
        {liao2024modeling}
\bibfield{author}{\bibinfo{person}{Zihan Liao}, \bibinfo{person}{Xiaodong Wu}, \bibinfo{person}{Shuo Shang}, \bibinfo{person}{Jun Wang}, {and} \bibinfo{person}{Wei Zhang}.} \bibinfo{year}{2024}\natexlab{}.
\newblock \showarticletitle{Modeling dynamic item tendency bias in sequential recommendation with causal intervention}.
\newblock \bibinfo{journal}{\emph{IEEE Transactions on Knowledge and Data Engineering}} (\bibinfo{year}{2024}).
\newblock


\bibitem[Liu et~al\mbox{.}(2024)]%
        {liu2024mamba4rec}
\bibfield{author}{\bibinfo{person}{Chengkai Liu}, \bibinfo{person}{Jianghao Lin}, \bibinfo{person}{Jianling Wang}, \bibinfo{person}{Hanzhou Liu}, {and} \bibinfo{person}{James Caverlee}.} \bibinfo{year}{2024}\natexlab{}.
\newblock \showarticletitle{Mamba4Rec: Towards Efficient Sequential Recommendation with Selective State Space Models}.
\newblock \bibinfo{journal}{\emph{CoRR}} (\bibinfo{year}{2024}).
\newblock


\bibitem[Liu et~al\mbox{.}(2025)]%
        {liu2025csrec}
\bibfield{author}{\bibinfo{person}{Xiaoyu Liu}, \bibinfo{person}{Jiaxin Yuan}, \bibinfo{person}{Yuhang Zhou}, \bibinfo{person}{Jingling Li}, \bibinfo{person}{Furong Huang}, {and} \bibinfo{person}{Wei Ai}.} \bibinfo{year}{2025}\natexlab{}.
\newblock \showarticletitle{CSRec: Rethinking Sequential Recommendation from A Causal Perspective.}. In \bibinfo{booktitle}{\emph{Proceedings of the 48th International ACM SIGIR Conference on Research and Development in Information Retrieval}}. \bibinfo{pages}{1562--1571}.
\newblock


\bibitem[Ma et~al\mbox{.}(2024)]%
        {pdrec}
\bibfield{author}{\bibinfo{person}{Haokai Ma}, \bibinfo{person}{Ruobing Xie}, \bibinfo{person}{Lei Meng}, \bibinfo{person}{Xin Chen}, \bibinfo{person}{Xu Zhang}, \bibinfo{person}{Leyu Lin}, {and} \bibinfo{person}{Zhanhui Kang}.} \bibinfo{year}{2024}\natexlab{}.
\newblock \showarticletitle{Plug-in diffusion model for sequential recommendation}. In \bibinfo{booktitle}{\emph{Proceedings of the AAAI conference on artificial intelligence}}, Vol.~\bibinfo{volume}{38}. \bibinfo{pages}{8886--8894}.
\newblock


\bibitem[Ma and Yu(2023)]%
        {ma2023selection}
\bibfield{author}{\bibinfo{person}{Teng Ma} {and} \bibinfo{person}{Su Yu}.} \bibinfo{year}{2023}\natexlab{}.
\newblock \showarticletitle{De-Selection Bias Recommendation Algorithm Based on Propensity Score Estimation}.
\newblock \bibinfo{journal}{\emph{Applied Sciences}} \bibinfo{volume}{13}, \bibinfo{number}{14} (\bibinfo{year}{2023}), \bibinfo{pages}{8038}.
\newblock


\bibitem[Ovaisi et~al\mbox{.}(2020)]%
        {10.1145/3366423.3380255}
\bibfield{author}{\bibinfo{person}{Zohreh Ovaisi}, \bibinfo{person}{Ragib Ahsan}, \bibinfo{person}{Yifan Zhang}, \bibinfo{person}{Kathryn Vasilaky}, {and} \bibinfo{person}{Elena Zheleva}.} \bibinfo{year}{2020}\natexlab{}.
\newblock \showarticletitle{Correcting for Selection Bias in Learning-to-rank Systems}. In \bibinfo{booktitle}{\emph{Proceedings of The Web Conference 2020}} (Taipei, Taiwan) \emph{(\bibinfo{series}{WWW '20})}. \bibinfo{publisher}{Association for Computing Machinery}, \bibinfo{address}{New York, NY, USA}, \bibinfo{pages}{1863–1873}.
\newblock
\showISBNx{9781450370233}
\href{https://doi.org/10.1145/3366423.3380255}{doi:\nolinkurl{10.1145/3366423.3380255}}


\bibitem[Pegoraro~Santana et~al\mbox{.}(2020)]%
        {music4all}
\bibfield{author}{\bibinfo{person}{Igor~André Pegoraro~Santana}, \bibinfo{person}{Fabio Pinhelli}, \bibinfo{person}{Juliano Donini}, \bibinfo{person}{Leonardo Catharin}, \bibinfo{person}{Rafael~Biazus Mangolin}, \bibinfo{person}{Yandre Maldonado e~Gomes da Costa}, \bibinfo{person}{Valéria Delisandra~Feltrim}, {and} \bibinfo{person}{Marcos~Aurélio Domingues}.} \bibinfo{year}{2020}\natexlab{}.
\newblock \showarticletitle{Music4All: A New Music Database and Its Applications}. In \bibinfo{booktitle}{\emph{2020 International Conference on Systems, Signals and Image Processing (IWSSIP)}}. \bibinfo{pages}{399--404}.
\newblock
\href{https://doi.org/10.1109/IWSSIP48289.2020.9145170}{doi:\nolinkurl{10.1109/IWSSIP48289.2020.9145170}}


\bibitem[Qi et~al\mbox{.}(2022)]%
        {10.1145/3477495.3532046}
\bibfield{author}{\bibinfo{person}{Tao Qi}, \bibinfo{person}{Fangzhao Wu}, \bibinfo{person}{Chuhan Wu}, \bibinfo{person}{Peijie Sun}, \bibinfo{person}{Le Wu}, \bibinfo{person}{Xiting Wang}, \bibinfo{person}{Yongfeng Huang}, {and} \bibinfo{person}{Xing Xie}.} \bibinfo{year}{2022}\natexlab{}.
\newblock \showarticletitle{ProFairRec: Provider Fairness-aware News Recommendation}. In \bibinfo{booktitle}{\emph{Proceedings of the 45th International ACM SIGIR Conference on Research and Development in Information Retrieval}} (Madrid, Spain) \emph{(\bibinfo{series}{SIGIR '22})}. \bibinfo{publisher}{Association for Computing Machinery}, \bibinfo{address}{New York, NY, USA}, \bibinfo{pages}{1164–1173}.
\newblock
\showISBNx{9781450387323}
\href{https://doi.org/10.1145/3477495.3532046}{doi:\nolinkurl{10.1145/3477495.3532046}}


\bibitem[Shi et~al\mbox{.}(2023)]%
        {10.1016/j.eswa.2022.118932}
\bibfield{author}{\bibinfo{person}{Lei Shi}, \bibinfo{person}{Shuqing Li}, \bibinfo{person}{Xiaowei Ding}, {and} \bibinfo{person}{Zhan Bu}.} \bibinfo{year}{2023}\natexlab{}.
\newblock \showarticletitle{Selection bias mitigation in recommender system using uninteresting items based on temporal visibility}.
\newblock \bibinfo{journal}{\emph{Expert Syst. Appl.}} \bibinfo{volume}{213}, \bibinfo{number}{PA} (\bibinfo{date}{March} \bibinfo{year}{2023}), \bibinfo{numpages}{11}~pages.
\newblock
\showISSN{0957-4174}
\href{https://doi.org/10.1016/j.eswa.2022.118932}{doi:\nolinkurl{10.1016/j.eswa.2022.118932}}


\bibitem[Sun et~al\mbox{.}(2019)]%
        {sun2019bert4rec}
\bibfield{author}{\bibinfo{person}{Fei Sun}, \bibinfo{person}{Jun Liu}, \bibinfo{person}{Jian Wu}, \bibinfo{person}{Changhua Pei}, \bibinfo{person}{Xiao Lin}, \bibinfo{person}{Wenwu Ou}, {and} \bibinfo{person}{Peng Jiang}.} \bibinfo{year}{2019}\natexlab{}.
\newblock \showarticletitle{BERT4Rec: Sequential recommendation with bidirectional encoder representations from transformer}. In \bibinfo{booktitle}{\emph{Proceedings of the 28th ACM international conference on information and knowledge management}}. \bibinfo{pages}{1441--1450}.
\newblock


\bibitem[Vaswani et~al\mbox{.}(2017)]%
        {vaswani2017attention}
\bibfield{author}{\bibinfo{person}{Ashish Vaswani}, \bibinfo{person}{Noam Shazeer}, \bibinfo{person}{Niki Parmar}, \bibinfo{person}{Jakob Uszkoreit}, \bibinfo{person}{Llion Jones}, \bibinfo{person}{Aidan~N Gomez}, \bibinfo{person}{{\L}ukasz Kaiser}, {and} \bibinfo{person}{Illia Polosukhin}.} \bibinfo{year}{2017}\natexlab{}.
\newblock \showarticletitle{Attention is all you need}.
\newblock \bibinfo{journal}{\emph{Advances in neural information processing systems}}  \bibinfo{volume}{30} (\bibinfo{year}{2017}).
\newblock


\bibitem[Wan and McAuley(2018)]%
        {goodreads1}
\bibfield{author}{\bibinfo{person}{Mengting Wan} {and} \bibinfo{person}{Julian~J. McAuley}.} \bibinfo{year}{2018}\natexlab{}.
\newblock \showarticletitle{Item recommendation on monotonic behavior chains}. In \bibinfo{booktitle}{\emph{Proceedings of the 12th {ACM} Conference on Recommender Systems, RecSys 2018, Vancouver, BC, Canada, October 2-7, 2018}}, \bibfield{editor}{\bibinfo{person}{Sole Pera}, \bibinfo{person}{Michael~D. Ekstrand}, \bibinfo{person}{Xavier Amatriain}, {and} \bibinfo{person}{John O'Donovan}} (Eds.). \bibinfo{publisher}{{ACM}}, \bibinfo{pages}{86--94}.
\newblock
\href{https://doi.org/10.1145/3240323.3240369}{doi:\nolinkurl{10.1145/3240323.3240369}}


\bibitem[Wan et~al\mbox{.}(2019)]%
        {goodreads2}
\bibfield{author}{\bibinfo{person}{Mengting Wan}, \bibinfo{person}{Rishabh Misra}, \bibinfo{person}{Ndapa Nakashole}, {and} \bibinfo{person}{Julian~J. McAuley}.} \bibinfo{year}{2019}\natexlab{}.
\newblock \showarticletitle{Fine-Grained Spoiler Detection from Large-Scale Review Corpora}. In \bibinfo{booktitle}{\emph{Proceedings of the 57th Conference of the Association for Computational Linguistics, {ACL} 2019, Florence, Italy, July 28- August 2, 2019, Volume 1: Long Papers}}, \bibfield{editor}{\bibinfo{person}{Anna Korhonen}, \bibinfo{person}{David~R. Traum}, {and} \bibinfo{person}{Llu{\'{\i}}s M{\`{a}}rquez}} (Eds.). \bibinfo{publisher}{Association for Computational Linguistics}, \bibinfo{pages}{2605--2610}.
\newblock
\href{https://doi.org/10.18653/V1/P19-1248}{doi:\nolinkurl{10.18653/V1/P19-1248}}


\bibitem[Wang et~al\mbox{.}(2024b)]%
        {wang2024causally}
\bibfield{author}{\bibinfo{person}{Lei Wang}, \bibinfo{person}{Chen Ma}, \bibinfo{person}{Xian Wu}, \bibinfo{person}{Zhaopeng Qiu}, \bibinfo{person}{Yefeng Zheng}, {and} \bibinfo{person}{Xu Chen}.} \bibinfo{year}{2024}\natexlab{b}.
\newblock \showarticletitle{Causally debiased time-aware recommendation}. In \bibinfo{booktitle}{\emph{Proceedings of the ACM Web Conference 2024}}. \bibinfo{pages}{3331--3342}.
\newblock


\bibitem[Wang et~al\mbox{.}(2023)]%
        {tdiffrec}
\bibfield{author}{\bibinfo{person}{Wenjie Wang}, \bibinfo{person}{Yiyan Xu}, \bibinfo{person}{Fuli Feng}, \bibinfo{person}{Xinyu Lin}, \bibinfo{person}{Xiangnan He}, {and} \bibinfo{person}{Tat-Seng Chua}.} \bibinfo{year}{2023}\natexlab{}.
\newblock \showarticletitle{Diffusion recommender model}. In \bibinfo{booktitle}{\emph{Proceedings of the 46th international ACM SIGIR conference on research and development in information retrieval}}. \bibinfo{pages}{832--841}.
\newblock


\bibitem[Wang et~al\mbox{.}(2021)]%
        {10.1145/3437963.3441799}
\bibfield{author}{\bibinfo{person}{Xiaojie Wang}, \bibinfo{person}{Rui Zhang}, \bibinfo{person}{Yu Sun}, {and} \bibinfo{person}{Jianzhong Qi}.} \bibinfo{year}{2021}\natexlab{}.
\newblock \showarticletitle{Combating Selection Biases in Recommender Systems with a Few Unbiased Ratings}. In \bibinfo{booktitle}{\emph{Proceedings of the 14th ACM International Conference on Web Search and Data Mining}} (Virtual Event, Israel) \emph{(\bibinfo{series}{WSDM '21})}. \bibinfo{publisher}{Association for Computing Machinery}, \bibinfo{address}{New York, NY, USA}, \bibinfo{pages}{427–435}.
\newblock
\showISBNx{9781450382977}
\href{https://doi.org/10.1145/3437963.3441799}{doi:\nolinkurl{10.1145/3437963.3441799}}


\bibitem[Wang et~al\mbox{.}(2024a)]%
        {10.1007/978-981-97-2262-4_13}
\bibfield{author}{\bibinfo{person}{Yu Wang}, \bibinfo{person}{Zhiwei Liu}, \bibinfo{person}{Liangwei Yang}, {and} \bibinfo{person}{Philip~S. Yu}.} \bibinfo{year}{2024}\natexlab{a}.
\newblock \showarticletitle{Conditional Denoising Diffusion for\&nbsp;Sequential Recommendation}. In \bibinfo{booktitle}{\emph{Advances in Knowledge Discovery and Data Mining: 28th Pacific-Asia Conference on Knowledge Discovery and Data Mining, PAKDD 2024, Taipei, Taiwan, May 7–10, 2024, Proceedings, Part V}} (Taipei, Taiwan). \bibinfo{publisher}{Springer-Verlag}, \bibinfo{address}{Berlin, Heidelberg}, \bibinfo{pages}{156–169}.
\newblock
\showISBNx{978-981-97-2264-8}
\href{https://doi.org/10.1007/978-981-97-2262-4_13}{doi:\nolinkurl{10.1007/978-981-97-2262-4_13}}


\bibitem[Wang et~al\mbox{.}(2022b)]%
        {10.1145/3511808.3557268}
\bibfield{author}{\bibinfo{person}{Yu Wang}, \bibinfo{person}{Hengrui Zhang}, \bibinfo{person}{Zhiwei Liu}, \bibinfo{person}{Liangwei Yang}, {and} \bibinfo{person}{Philip~S. Yu}.} \bibinfo{year}{2022}\natexlab{b}.
\newblock \showarticletitle{ContrastVAE: Contrastive Variational AutoEncoder for Sequential Recommendation}. In \bibinfo{booktitle}{\emph{Proceedings of the 31st ACM International Conference on Information \& Knowledge Management}} (Atlanta, GA, USA) \emph{(\bibinfo{series}{CIKM '22})}. \bibinfo{publisher}{Association for Computing Machinery}, \bibinfo{address}{New York, NY, USA}, \bibinfo{pages}{2056–2066}.
\newblock
\showISBNx{9781450392365}
\href{https://doi.org/10.1145/3511808.3557268}{doi:\nolinkurl{10.1145/3511808.3557268}}


\bibitem[Wang et~al\mbox{.}(2022a)]%
        {10.1145/3485447.3512092}
\bibfield{author}{\bibinfo{person}{Zhenlei Wang}, \bibinfo{person}{Shiqi Shen}, \bibinfo{person}{Zhipeng Wang}, \bibinfo{person}{Bo Chen}, \bibinfo{person}{Xu Chen}, {and} \bibinfo{person}{Ji-Rong Wen}.} \bibinfo{year}{2022}\natexlab{a}.
\newblock \showarticletitle{Unbiased Sequential Recommendation with Latent Confounders}. In \bibinfo{booktitle}{\emph{Proceedings of the ACM Web Conference 2022}} (Virtual Event, Lyon, France) \emph{(\bibinfo{series}{WWW '22})}. \bibinfo{publisher}{Association for Computing Machinery}, \bibinfo{address}{New York, NY, USA}, \bibinfo{pages}{2195–2204}.
\newblock
\showISBNx{9781450390965}
\href{https://doi.org/10.1145/3485447.3512092}{doi:\nolinkurl{10.1145/3485447.3512092}}


\bibitem[Xie et~al\mbox{.}(2021)]%
        {10.1145/3442381.3449873}
\bibfield{author}{\bibinfo{person}{Zhe Xie}, \bibinfo{person}{Chengxuan Liu}, \bibinfo{person}{Yichi Zhang}, \bibinfo{person}{Hongtao Lu}, \bibinfo{person}{Dong Wang}, {and} \bibinfo{person}{Yue Ding}.} \bibinfo{year}{2021}\natexlab{}.
\newblock \showarticletitle{Adversarial and Contrastive Variational Autoencoder for Sequential Recommendation}. In \bibinfo{booktitle}{\emph{Proceedings of the Web Conference 2021}} (Ljubljana, Slovenia) \emph{(\bibinfo{series}{WWW '21})}. \bibinfo{publisher}{Association for Computing Machinery}, \bibinfo{address}{New York, NY, USA}, \bibinfo{pages}{449–459}.
\newblock
\showISBNx{9781450383127}
\href{https://doi.org/10.1145/3442381.3449873}{doi:\nolinkurl{10.1145/3442381.3449873}}


\bibitem[Xu et~al\mbox{.}(2024)]%
        {xu2024rethinking}
\bibfield{author}{\bibinfo{person}{Wujiang Xu}, \bibinfo{person}{Qitian Wu}, \bibinfo{person}{Runzhong Wang}, \bibinfo{person}{Mingming Ha}, \bibinfo{person}{Qiongxu Ma}, \bibinfo{person}{Linxun Chen}, \bibinfo{person}{Bing Han}, {and} \bibinfo{person}{Junchi Yan}.} \bibinfo{year}{2024}\natexlab{}.
\newblock \showarticletitle{Rethinking cross-domain sequential recommendation under open-world assumptions}. In \bibinfo{booktitle}{\emph{Proceedings of the ACM Web Conference 2024}}. \bibinfo{pages}{3173--3184}.
\newblock


\bibitem[Yang et~al\mbox{.}(2024)]%
        {yang2024debiasing}
\bibfield{author}{\bibinfo{person}{Jiyuan Yang}, \bibinfo{person}{Yue Ding}, \bibinfo{person}{Yidan Wang}, \bibinfo{person}{Pengjie Ren}, \bibinfo{person}{Zhumin Chen}, \bibinfo{person}{Fei Cai}, \bibinfo{person}{Jun Ma}, \bibinfo{person}{Rui Zhang}, \bibinfo{person}{Zhaochun Ren}, {and} \bibinfo{person}{Xin Xin}.} \bibinfo{year}{2024}\natexlab{}.
\newblock \showarticletitle{Debiasing sequential recommenders through distributionally robust optimization over system exposure}. In \bibinfo{booktitle}{\emph{Proceedings of the 17th ACM International Conference on Web Search and Data Mining}}. \bibinfo{pages}{882--890}.
\newblock


\bibitem[Yu et~al\mbox{.}(2025)]%
        {11150713}
\bibfield{author}{\bibinfo{person}{Dianer Yu}, \bibinfo{person}{Qian Li}, \bibinfo{person}{Sirui Huang}, \bibinfo{person}{Jie Cao}, {and} \bibinfo{person}{Guandong Xu}.} \bibinfo{year}{2025}\natexlab{}.
\newblock \showarticletitle{Large Language Models meet Causal Inference: Semantic-Rich Dual Propensity Score for Sequential Recommendation}.
\newblock \bibinfo{journal}{\emph{IEEE Transactions on Knowledge and Data Engineering}} (\bibinfo{year}{2025}), \bibinfo{pages}{1--12}.
\newblock
\href{https://doi.org/10.1109/TKDE.2025.3606149}{doi:\nolinkurl{10.1109/TKDE.2025.3606149}}


\bibitem[Yu et~al\mbox{.}(2023)]%
        {10.1016/j.neucom.2023.01.089}
\bibfield{author}{\bibinfo{person}{Dianer Yu}, \bibinfo{person}{Qian Li}, \bibinfo{person}{Xiangmeng Wang}, {and} \bibinfo{person}{Guandong Xu}.} \bibinfo{year}{2023}\natexlab{}.
\newblock \showarticletitle{Deconfounded recommendation via causal intervention}.
\newblock \bibinfo{journal}{\emph{Neurocomput.}} \bibinfo{volume}{529}, \bibinfo{number}{C} (\bibinfo{date}{April} \bibinfo{year}{2023}), \bibinfo{pages}{128–139}.
\newblock
\showISSN{0925-2312}
\href{https://doi.org/10.1016/j.neucom.2023.01.089}{doi:\nolinkurl{10.1016/j.neucom.2023.01.089}}


\bibitem[Yue et~al\mbox{.}(2024)]%
        {10.1145/3616855.3635760}
\bibfield{author}{\bibinfo{person}{Zhenrui Yue}, \bibinfo{person}{Yueqi Wang}, \bibinfo{person}{Zhankui He}, \bibinfo{person}{Huimin Zeng}, \bibinfo{person}{Julian Mcauley}, {and} \bibinfo{person}{Dong Wang}.} \bibinfo{year}{2024}\natexlab{}.
\newblock \showarticletitle{Linear Recurrent Units for Sequential Recommendation}. In \bibinfo{booktitle}{\emph{Proceedings of the 17th ACM International Conference on Web Search and Data Mining}} (Merida, Mexico) \emph{(\bibinfo{series}{WSDM '24})}. \bibinfo{publisher}{Association for Computing Machinery}, \bibinfo{address}{New York, NY, USA}, \bibinfo{pages}{930–938}.
\newblock
\showISBNx{9798400703713}
\href{https://doi.org/10.1145/3616855.3635760}{doi:\nolinkurl{10.1145/3616855.3635760}}


\bibitem[Zhang et~al\mbox{.}(2024)]%
        {zhang2024uncovering}
\bibfield{author}{\bibinfo{person}{Honglei Zhang}, \bibinfo{person}{Shuyi Wang}, \bibinfo{person}{Haoxuan Li}, \bibinfo{person}{Chunyuan Zheng}, \bibinfo{person}{Xu Chen}, \bibinfo{person}{Li Liu}, \bibinfo{person}{Shanshan Luo}, {and} \bibinfo{person}{Peng Wu}.} \bibinfo{year}{2024}\natexlab{}.
\newblock \showarticletitle{Uncovering the propensity identification problem in debiased recommendations}. In \bibinfo{booktitle}{\emph{2024 IEEE 40th International Conference on Data Engineering (ICDE)}}. IEEE, \bibinfo{pages}{653--666}.
\newblock


\bibitem[Zhao et~al\mbox{.}(2023)]%
        {10.1109/TKDE.2022.3218994}
\bibfield{author}{\bibinfo{person}{Zihao Zhao}, \bibinfo{person}{Jiawei Chen}, \bibinfo{person}{Sheng Zhou}, \bibinfo{person}{Xiangnan He}, \bibinfo{person}{Xuezhi Cao}, \bibinfo{person}{Fuzheng Zhang}, {and} \bibinfo{person}{Wei Wu}.} \bibinfo{year}{2023}\natexlab{}.
\newblock \showarticletitle{Popularity Bias is not Always Evil: Disentangling Benign and Harmful Bias for Recommendation}.
\newblock \bibinfo{journal}{\emph{IEEE Trans. on Knowl. and Data Eng.}} \bibinfo{volume}{35}, \bibinfo{number}{10} (\bibinfo{date}{Oct.} \bibinfo{year}{2023}), \bibinfo{pages}{9920–9931}.
\newblock
\showISSN{1041-4347}
\href{https://doi.org/10.1109/TKDE.2022.3218994}{doi:\nolinkurl{10.1109/TKDE.2022.3218994}}


\bibitem[Zhao et~al\mbox{.}(2025)]%
        {10.1145/3726302.3730005}
\bibfield{author}{\bibinfo{person}{Ziqi Zhao}, \bibinfo{person}{Zhaochun Ren}, \bibinfo{person}{Jiyuan Yang}, \bibinfo{person}{Zuming Yan}, \bibinfo{person}{Zihan Wang}, \bibinfo{person}{Liu Yang}, \bibinfo{person}{Pengjie Ren}, \bibinfo{person}{Zhumin Chen}, \bibinfo{person}{Maarten de Rijke}, {and} \bibinfo{person}{Xin Xin}.} \bibinfo{year}{2025}\natexlab{}.
\newblock \showarticletitle{Improving Sequential Recommenders through Counterfactual Augmentation of System Exposure}. In \bibinfo{booktitle}{\emph{Proceedings of the 48th International ACM SIGIR Conference on Research and Development in Information Retrieval}} (Padua, Italy) \emph{(\bibinfo{series}{SIGIR '25})}. \bibinfo{publisher}{Association for Computing Machinery}, \bibinfo{address}{New York, NY, USA}, \bibinfo{pages}{1508–1518}.
\newblock
\showISBNx{9798400715921}
\href{https://doi.org/10.1145/3726302.3730005}{doi:\nolinkurl{10.1145/3726302.3730005}}


\bibitem[Zhu et~al\mbox{.}(2020)]%
        {10.1145/3397271.3401177}
\bibfield{author}{\bibinfo{person}{Ziwei Zhu}, \bibinfo{person}{Jianling Wang}, {and} \bibinfo{person}{James Caverlee}.} \bibinfo{year}{2020}\natexlab{}.
\newblock \showarticletitle{Measuring and Mitigating Item Under-Recommendation Bias in Personalized Ranking Systems}. In \bibinfo{booktitle}{\emph{Proceedings of the 43rd International ACM SIGIR Conference on Research and Development in Information Retrieval}} (Virtual Event, China) \emph{(\bibinfo{series}{SIGIR '20})}. \bibinfo{publisher}{Association for Computing Machinery}, \bibinfo{address}{New York, NY, USA}, \bibinfo{pages}{449–458}.
\newblock
\showISBNx{9781450380164}
\href{https://doi.org/10.1145/3397271.3401177}{doi:\nolinkurl{10.1145/3397271.3401177}}


\end{thebibliography}
\end{document}